\documentstyle[12pt]{article}
\newcommand{\FF}{{\cal F}}
\newcommand{\LL}{{\cal L}}
\newcommand{\OO}{{\cal O}}
\newcommand{\rvac}{|0\rangle}
\newcommand{\lvac}{\langle 0|}
\renewcommand{\=}{&=&} %seems not to work in footnotes
\newcommand{\ie}{{{\em i.e.},\ }}
\newcommand{\eg}{{{\em e.g.},\ }}
\renewcommand{\bar}{\overline}
\renewcommand{\a}{\alpha}
\renewcommand{\b}{\beta}
\renewcommand{\d}{\delta}
\newcommand{\g}{\gamma}
\renewcommand{\l}{\lambda}
\newcommand{\m}{\mu}
\newcommand{\n}{\nu}
\renewcommand{\o}{\omega}
\newcommand{\s}{\sigma}
\renewcommand{\t}{\theta}
\newcommand{\tb}{{\bar\theta}}
\newcommand{\aad}{{\a\ad}}
\newcommand{\ad}{{\dot{\alpha}}}
\newcommand{\bd}{{\dot{\beta}}}
\newcommand{\pa}{\partial}
\newcommand{\th}{\widehat\theta}
\newcommand{\yh}{\widehat{y}}
\newcommand{\tbh}{\widehat{\bar\theta}}

\newcommand{\gd}{{\dot{\gamma}}}
\newcommand{\qh}{{\widehat Q}}
\newcommand{\qbh}{{\widehat{\bar Q}}}
\newcommand{\dh}{{\widehat D}}
\newcommand{\dbh}{{\widehat{\bar D}}}
\newcommand{\qb}{{\bar Q}}
\newcommand{\bb}{{\beta\kern-.41em \beta\kern-.41em \beta}}
\newcommand{\Tr}{{\rm Tr \,}}
\newcommand{\xx}{{\bf x}}

\newcommand{\id}{{\mathchoice {\rm 1\mskip-4mu l} {\rm 1\mskip-4mu l}
{\rm 1\mskip-4.5mu l} {\rm 1\mskip-5mu l}}}
\newcommand{\real}{{{\rm I} \kern -.19em {\rm R}}}
\newcommand{\y}{(y)}
\newcommand{\pad}[2]{{\displaystyle{\frac{\partial #1}{\partial #2}}}}
\newcommand{\mn}{{\mu\nu}}

\newcommand{\intq}{{\int d^4 \! x \,\,}}
\newcommand{\half}{\frac 1 2}
\newcommand{\intt}{{\int d^3 \! {\bf x} \,\,}}
\renewcommand{\ss}{\sigma\kern-.54em \sigma}
\newcommand{\be}{\begin{equation}}
\newcommand{\ee}{\end{equation}}
\newcommand{\beq}{\begin{eqnarray}}
\newcommand{\eeq}{\end{eqnarray}}
\newcommand{\ba}{\begin{array}}
\newcommand{\ea}{\end{array}}
\newcommand{\equ}[1]{(\ref{#1})}
\makeatletter
\@addtoreset{equation}{section}
\makeatother

\catcode`\@=11
\def\marginnote#1{}
\def\@eqnlabel{}
\def\@vacuum{}
\def\titlepage{\@restonecolfalse\if@twocolumn\@restonecoltrue\onecolumn
  \else \newpage \fi \thispagestyle{empty}\c@page\z@
  \def\thefootnote{\fnsymbol{footnote}} }
\def\endtitlepage{\if@restonecol\twocolumn \else \fi
  \def\thefootnote{\arabic{footnote}}
  \setcounter{footnote}{0}} %\c@footnote\z@ }
\catcode`@=12
\relax
\newcommand{\note}[1]{$\bullet\kern-.3em\bullet$[{\bf{#1}}]}
\newcommand{\eqnote}[1]{\bullet\kern-.3em\bullet[{\bf{#1}}]}

\begin{document}
\begin{titlepage}
\begin{flushright}
NEIP--98--006 \\
hep--ph/9807403 \\
July 1998
\end{flushright}
\vskip 0.25in
\begin{center}{\Large\bf
REALIZATIONS OF THERMAL \\[2mm] SUPERSYMMETRY}
\vskip .2in
{\bf Jean-Pierre Derendinger and Claudio
Lucchesi\footnote{Supported by the Swiss
National Science Foundation.} }
\vskip .1in
Institut de Physique \\
Universit\'e de Neuch\^atel \\
CH--2000 Neuch\^atel, Switzerland
\end{center}
\vspace{1.7cm}
\begin{center}
{\bf Abstract}
\end{center}
\begin{quote}
{We investigate realizations of supersymmetry at finite temperature
in terms of thermal superfields, in a thermally constrained superspace:
the Grassmann coordinates are promoted to be time-dependent and
antiperiodic, with a period given by the inverse temperature.
This approach allows to formulate a Kubo-Martin-Schwinger (KMS)
condition at the level of thermal superfield propagators. The latter
is proven directly in thermal superspace, and is shown to imply the correct
(bosonic and fermionic) KMS conditions for the component fields.
In thermal superspace, we formulate thermal covariant derivatives and
supercharges and derive the thermal super-Poincar\'e algebra.
Finally, we briefly investigate field realizations of this thermal
supersymmetry algebra, focussing on the Wess-Zumino model.
The thermal superspace formalism is used to characterize the breaking
of global supersymmetry at finite temperature.}
\end{quote}

\end{titlepage}
\setcounter{footnote}{0}
\setcounter{page}{0}
\newpage

\section{Introduction }\label{intro}

The most popular extension of the Standard Model of strong and electroweak
interactions has a broken supersymmetry. The breaking of supersymmetry
is  in several aspects different from the breaking of, for instance,
the $SU(2)_L\times U(1)_Y$ gauge symmetry. Firstly, it is not spontaneous,
or it is spontaneous only in a more fundamental theory like supergravity or
superstrings acting at much shorter distances.
Secondly, in contrast with the common behaviour of spontaneously broken
gauge symmetries, supersymmetry is not restored at high temperature. This
fact has conceptual and technical implications when studying
phase transitions in the
minimal supersymmetric standard model or, more generally, cosmology of
supersymmetric field theories. One cannot expect that the cooling  of the
expanding Universe has triggered supersymmetry breaking.

The fate of supersymmetry in thermal environments
has been studied with various
motivations and purposes in a relatively small number of references
\cite{dk}--\cite{bo}, scattered over the last
twenty years. The behaviour of supersymmetry
at finite temperature is somewhat but not entirely similar to the one of
Lorentz symmetry. In a thermal bath, a relativistic field theory can be
formulated in a covariant way \cite{weldon}, even if the notion of
temperature is certainly not Lorentz invariant since it refers to
a particular frame. Poincar\'e symmetry {\it acts} then in a well
defined way on the observables of the theory. Supersymmetry adds a
new aspect: it exchanges bosons and fermions. At zero temperature,
superspace takes care of the difference in statistics by introducing
Grassmann anticommuting coordinates, which turn bosonic commutation
relations into fermionic anticommutators. At finite temperature,
fermion and boson statistics involve in addition
the appropriate statistical weight in field theory Green's functions.

Our first goal in Sections 2 and 3 of
the present paper is to develop a superspace
covariant formalism for the Green's functions of a supersymmetric field
theory at finite temperature.
We formulate the restrictions that thermal
effects impose on superspace, in terms of Kubo-Martin-Schwinger (KMS)
relations \cite{kms} on Green's functions. In sections 4 and 5, we then use
this ``thermal superspace" formalism
to derive thermal covariant derivatives, thermal supercharges, as well as the
thermal covariantizations of the translation and Lorentz generators,
in complete analogy with the zero-temperature case.
These operators generate the usual super-Poincar\'e algebra, acting
locally on the thermal version of superspace. Notice that the existence of
this algebra is not in contradiction with the expected breaking
of supersymmetry in finite temperature field theories, which is due to
periodicity (for bosons) and antiperiodicity (for fermions) relations in
the (complex) time direction. These characterizations are of
global character, and the super-Poincar\'e algebra is not sensitive
to the field's global periodicity properties. But the existence of
a super-Poincar\'e algebra in superspace implies neither the existence
of field representations nor the existence of invariant actions.

At a technical level, the present approach rests mainly on the notion of
thermal superspace, the properties of which can be motivated through the
following, heuristic argument. Consider first supersymm{e}try at zero
temperature. The  supersymmetry algebra relates supersymmetries $Q_\a$, ${\bar
Q}_\ad$ to translations $P_\m$ through $\{Q_\a,{\bar Q}_\bd\}=2 (\s^\m)_{\a\bd}
P_\m$. As a consequence, supersymmetries  can be seen as ``square roots" of
translations, which act like the derivative, $P_\m=-i\pa_\m$. One could thus
try to represent supersymmetries as generalized translations,
which would act through derivatives only, in analogy to translations. This
is however not possible in usual space-time, but requires an
enlarged space including the Grassmann variables that are translated by
the supersymmetry generators. Such an enlarged space is provided by superspace,
which consists, in addition to the usual space-time coordinates, of
grassmannian objects denoted by $\t$ and $\tb$. A point $X$ in superspace has
therefore coordinates $X = (x^\m,\t^\a,\tb^\ad)$,
and since at zero temperature the parameters of supersymmetry transformations
are space-time constant, the zero-temperature superspace coordinates  $\t^\a$
and $\tb^\ad$  are space-time constants as well.

However, as was noted in  {\cite{ggs} (see also \cite{boyan})}, if one wants to
respect the correct  KMS boundary conditions, one {\it cannot} make use of {\it
constant}  parameters in supersymmetry transformations rules at finite
temperature : the supersymmetry parameters must be {\it time-dependent} and
{\it antiperiodic}  in imaginary time on the interval $[0,\b]$, where $\b =
1/T$ denotes the inverse temperature{\footnote{For a discussion of the
difficulties in realizing the correct boundary conditions for thermal fields
with constant supersymmetry parameters, see \cite{fujikawa}.}}.
Adapting straightforwardly the zero-temperature argument above, it is  natural
to require that the variables which are translated by the effect of the thermal
supersymmetry transformation bear the same characteristics as the thermal
supersymmetry parameters. From this we conclude that  the construction of a
thermal superspace requires that the Grassmann parameters get promoted to be
{\it time-dependent} and {\it antiperiodic}  in imaginary time on the interval
$[0,\b]$. A point in thermal superspace has therefore coordinates
\be
\widehat X = \Bigl(x^\m,{\widehat\t}^\a(t),{\widehat\tb}^\ad(t)\Bigr)\, ,
\label{thsu}
\ee
where a ``hat" is used to denote thermal quantities, and ${\widehat\t}^\a(t)$,
${\widehat\tb}^\ad(t)$ are subject to the {\it antiperiodicity conditions}
\be
{\widehat\t}^\a(t+i\b)=-{\widehat\t}^\a(t)\, ,\qquad
{\widehat\tb}^\ad(t+i\b)=-{\widehat\tb}^\ad(t)\, .
\label{antiper}
\ee
This argument is at the basis of the development of thermal superspace
in the following sections.

The last part of the paper considers field representations of
the thermal super-Poincar\'e algebra, focussing
on the simplest case of a chiral supermultiplet (Sections 6 and 7).
Thermal fields are characterized by their time/temperature
periodicity properties: bosonic fields are periodic,
while fermionic fields are antiperiodic. We apply our thermal superspace
formalism to the free Wess-Zumino model and briefly study the behaviour
of the model under thermal supersymmetry transformations, and the structure
of its breaking.
Starting from the $T=0$  Wess-Zumino model, we perform the Matsubara mode
expansion on the temperature circle of length $\beta$ and derive the
resulting three-dimensional action for thermal modes. We obtain the bosonic
and fermionic thermal mass spectra. Finally, we derive the variation
of the finite temperature model under thermal supersymmetry, using the
supercharges obtained in thermal superspace.

\section{Thermal formalism for bosons and fermions}\label{tfbf}

As supersymmetric theories include bosonic and fermionic fields,
we start this section by briefly reviewing some fundamentals of
thermal field theory for bosons and fermions.

We will mainly consider physical systems with free fields
immersed in a thermal bath at non-zero temperature $T$.
At the level of field averages, this means that we have to
replace the vacuum expectation value $\lvac...|0\rangle$ by the thermal
average $\langle ... \rangle_\bb$ (to be defined below). The thermal average
involves a weighted summation over all accessible eigenstates $|n\rangle$ of
the hamiltonian $H$,
\be
H |n\rangle = E_n |n\rangle\qquad n=0,1,\ldots,
\label{eshh}
\ee
the spectrum of which is  discrete and infinite. The lowest energy state
is $|0\rangle$, with energy $E_0$. We assume that the system does not exchange
particles with the heat bath, that is, we set the chemical potential to zero.
The relevant statistical partition function is the canonical one, given by the
trace of the Boltzmann weight,
\begin{equation}\label{pf}
Z(\b) = \Tr e^{-\beta H}\, ,
\end{equation}
where $\beta=1/T$ is the inverse temperature (we set the Boltzmann constant
$k_{\rm B}=1$), and the quantum-mechanical trace is
\be
\Tr \OO = \sum_n \langle n|\OO |n\rangle\, .
\label{trace}
\ee
The {\it thermal average} $\langle\ ...\rangle_\bb$ of an arbitrary field
operator $\OO$ is then defined as
\be
\langle \OO\rangle_\b \equiv {1\over Z(\b)}\ \Tr\left( e^{- \beta H} \OO
\right),
\label{limve}
\ee
with the usual Boltzmann weight factor and the normalization $\langle
\id\rangle_\bb =1$. We also assume that the lowest energy state has $E_0=0$
so that, in the zero-temperature limit,
$$
\langle \OO\rangle_\b \,\,\stackrel{\beta\rightarrow\infty}{\longrightarrow}
\,\, \langle0|\OO|0\rangle.
$$
This assumption is merely a normal ordering prescription.

Through the LSZ reduction formulae, a field theory can be defined by
the whole set of its Green's functions. At finite temperature, the Green's
functions are subject to important periodicity constraints in imaginary time,
known as the KMS (Kubo-Martin-Schwinger) conditions \cite{kms,tft}.
We now review the derivation of these conditions for bosonic and fermionic
fields, in view of later on discussing the thermal behaviour of superfields,
which contain both field statistics in their components.

Consider firstly a free real scalar field $\phi$ carrying no conserved
charges. The hamiltonian $H$ being the time evolution operator, the
field operator $\phi$ at $x=(t,\xx)$ (in the Heisenberg picture and
with $\hbar=c=1$) is
\begin{equation}\label{foh}
\phi(x)= \phi(t,{\xx})=e^{iHt}\phi(0,{\xx})e^{-iHt}\, ,
\end{equation}
with a time coordinate $x^0 =t$ which is allowed to be complex. We now define
the {\it $n$-point thermal Green's function} $G_{n\,C}$ to be the thermal
average of the
${\sf T}_{\!C}$-ordered product of the Heisenberg field,
\begin{equation}\label{tgc}
G_{n\,C}(x_1,...,x_n) =\langle {\sf T}_{\!C}\ \phi(x_1)
...\phi(x_n)\rangle_\bb\, .
\end{equation}
 The ``path-ordering" operation\footnote{Path-ordering is the thermal
generalization of the usual time-ordering of zero-temperature field theory. It
prescribes that the fields be arranged depending on the position of their
(complex) time  variables along a path $C$ taken in the complex time plane
\cite{tft}.} denoted by ${\sf T}_{\!C}$ is peculiar to the thermal case. For
any scalar field $\phi$, path ordering can be defined through
\begin{equation}\label{pord}
{\sf T}_{\!C}\ \phi(x_1)\phi(x_2)=\t_C(t_1-t_2)\phi(x_1)\phi(x_2)
+\t_C(t_2-t_1)\phi(x_2)\phi(x_1)\, ,
\end{equation}
where the path Heaviside function $\t_C$ is defined as $\t_C(t)\equiv
\t(\tau)$ for a path parametrized by $t=z(\tau)$, $\tau\in\real$. The thermal
path-ordered propagator then writes
\begin{equation}\label{orf}
D_{C}(x_1,x_2)=\t_C(t_1-t_2)D^>_{C}(x_1,x_2)+\t_C(t_2-t_1)D^<_{C}(x_1,x_2)\, ,
\end{equation}
where $D^>_{C}$, $D^<_{C}$ denote respectively the thermal bosonic two-point
functions
\begin{eqnarray}
D^>_{C}(x_1,x_2)\=\langle \phi(x_1)\phi(x_2)\rangle_\bb\, ,\label{adv}\\[2mm]
D^<_{C}(x_1,x_2)\=\langle\phi(x_2)\phi(x_1)\rangle_\bb\, .\label{ret}
\end{eqnarray}

The Boltzmann weight $e^{-\b H}$ can be interpreted as an
evolution operator in imaginary time. Indeed, rewriting the bosonic Heisenberg
field \equ{foh} for a translation in imaginary time by $t=i\beta$,
$\beta\in\real$, one gets
\be
e^{-\b H} \phi(t,\xx) e^{\b H}= \phi(t+i\b,\xx).
\label{bit}
\ee
Expressing \equ{adv} as
\be
D_{C}^> (x_1,x_2) = {1\over Z(\b)}\ \Tr \left[ e^{- \beta H}
\phi(x_1)\phi(x_2)\right],
\label{start}
\ee
using the cyclicity of the thermal trace (upon inserting $e^{\beta H}
e^{-\beta H} =1$) and the evolution in imaginary time \equ{bit}, one  deduces
the {\it bosonic KMS condition} \cite{kms,tft}. This condition relates
$D_{C}^>$ and $D_{C}^<$ through a translation in imaginary time,
\begin{equation}\label{kmsb}
D_{C}^> (t_1;\xx_1,t_2;\xx_2) = D_{C}^< (t_1+i\b;\xx_1,t_2;\xx_2)\, .
\end{equation}

A similar analysis can be performed for fermionic fields. Fermionic path
ordering differs from the bosonic case through a sign,
\begin{equation}\label{fpord}
{\sf T}_{\!C}\
\psi_a(x_1)\bar\psi_b(x_2)=\t_C(t_1-t_2)\psi_a(x_1)\bar\psi_b(x_2)
-\t_C(t_2-t_1)\bar\psi_b(x_2)\psi_a(x_1),
\end{equation}
with $a,b=1,...,4$ for Dirac (four-component) spinors. The thermal path-ordered
fermion propagator writes similarly to
\equ{orf},
\begin{equation}\label{forf}
S_{C\, ab}(x_1,x_2)
=\t_C(t_1-t_2)S^>_{C\, ab}(x_1,x_2)+\t_C(t_2-t_1)S^<_{C\, ab}(x_1,x_2),
\end{equation}
where $S^>_{C}$, $S^<_{C}$ are the thermal fermionic two-point functions
conventionally defined as
\begin{eqnarray}
S_{C\, {ab}}^{>}(x_1,x_2)\=\phantom{-}\langle
\psi_a(x_1)\bar\psi_b(x_2)\rangle_\bb,\label{advf}\\[2mm]
S _{C\, {ab}}^{<} (x_1,x_2)\=-\langle\bar\psi_b(x_2)\psi_a(x_1)\rangle_\bb
.\label{retf}
\end{eqnarray}
Following the same procedure as in the bosonic case, one derives the {\it
fermionic KMS condition}
\begin{equation}\label{kmsf}
S_{C\, {ab}}^{>} (t_1;\xx_1,t_2;\xx_2) =\ - S_{C\, {ab}}^{<}
(t_1+i\b;\xx_1,t_2;\xx_2)\, ,
\end{equation}
which differs from the bosonic one, eq. \equ{kmsb}, by a relative sign.

Superspace is usually formulated using two-component Weyl spinors,
$\psi_\alpha$ and $\bar\psi^\ad$, with respectively left-handed and
right-handed chirality. The relation with Dirac spinors is
$$
\psi_a = \left(\begin{array}{c} \psi_\alpha \\ \bar\psi^{\ad}
\end{array}\right).
$$
It is then useful to translate the KMS condition for Dirac spinors \equ{kmsf}
into KMS conditions for two-component spinors $\psi_\a$, $\bar\psi^\ad$.
Defining the thermal two-point functions $S_{C}^{>}$, $S_{C}^{<}$, for
two-component spinors as, \eg
\be
S_{C\,\a}^{>\ \ \dot\b}(x_1,x_2)=
\langle\psi_\a(x_1)\bar\psi^\bd(x_2)\rangle_\bb,
\qquad
S_{C\,\a}^{<\ \ \dot\b}(x_1,x_2)
=-\langle\bar\psi^\bd(x_2)\psi_\a(x_1)\rangle_\bb,
\label{afuu}
\ee
we derive from \equ{kmsf} a  {\it fermionic KMS condition for two-component
Majorana spinors}:
\be
S_{C\,\a}^{>\ \ \dot\b}(t_1;\xx_1,t_2;\xx_2)
= -S_{C\, \a}^{<\ \ \dot\b}(t_1+i\b;\xx_1,t_2;\xx_2).
\label{mmm1}
\ee
This is the only relation we shall need for practical purposes. But
similar relations can be derived for $S_{C\, \a\b}$,
$S_{C\ \ \b}^{\ \, \ad}$ and $S_{C\,}^{\ \, \ad\bd}$.

\section{Thermal formalism for superfields}\label{sectsf}

\subsection{Thermal propagators for superfields}\label{seccach}

At zero temperature, a superfield formulation of bosons and fermions is by
construction supersymmetric. The spinorial generators of supersymmetry
transformations -- the supercharges -- act like generalized derivatives on
superfields, which contain in their expansion the bosonic and fermionic
fields as components. And the supersymmetry transformation of superfields
encodes the transformations of its components.

As discussed previously, the KMS periodicity conditions provide an essential,
mandatory characterization of thermal effects at the level of Green's
functions. If superfield propagators can be defined in a thermal environment,
they should of course obey some form of KMS condition. And such a superfield
KMS condition should be able to reproduce the KMS boundary conditions for the
superfields' bosonic and fermionic components.

Let us start by recalling some simple facts of the zero-temperature
case. $T=0$ chiral superfields, noted $\phi$, and  $T=0$ antichiral
superfields, denoted by $\bar\phi$, are defined by the conditions
\be
\bar D_\ad \phi =0,\qquad\qquad D_\a{\bar\phi} =0,
\label{defcs}
\ee
where the $T=0$ covariant derivatives $D_\a$, $\bar D_\ad$ write
\be
D_\a= {\partial\over\partial \t^\a} - i\,\s^\m_{\ \aad}\tb^\ad\,\pa_\m\,,
\qquad\qquad
\bar D_\ad= {\partial\over\partial \tb^\ad}-i\,\t^\a\s^\m_{\ \aad}\,\pa_\m\,.
\label{covder}
\ee
Clearly, if $\phi$ is chiral, $\bar\phi=\phi^\dagger$ is antichiral.
A point in (zero-temperature) superspace is defined by the space-time
coordinates $x^\m$ and by anticommuting (Grassmann) spinor coordinates $\t^\a$
and $\tb^\ad$, which are space-time {\it constants}. It can be equivalently
defined by chiral coordinates $(y^\mu,\t^\a,\tb^\ad)$,
\beq\label{y}
y^\mu = x^\mu - i\t^\a\sigma^\mu_{\a\ad}\tb^\ad, \qquad
\bar D_\ad y^\mu=0,
\eeq
or by antichiral coordinates $(\bar y^\mu,\t^\a,\tb^\ad)$,
\beq
\label{ybar}
\bar y^\mu = x^\mu + i\t^\a\sigma^\mu_{\a\ad}\tb^\ad, \qquad
 D_\a y^\mu=0.
\eeq
In these variables, it follows from their defining equations \equ{defcs}
that the expansions of chiral and antichiral superfields are simply
\beq
\phi(y,\t)\= z\y+\sqrt{2}\,\t\,\psi\y-\t\t f\y\, ,\label{chsexp}\\[2mm]
{\bar\phi}({\bar y},\tb)\={\bar z}({\bar y})+\sqrt{2}\,\tb\,\bar\psi({\bar
y})-\tb\tb
{\bar f}({\bar y})\, .\label{xexpa}
\eeq
The components of the superfields $\phi$ and $\bar\phi$ form a chiral
multiplet, which contains two complex scalar fields $z$ and $f$ and a Majorana
spinor\footnote{The spinorial component of the chiral superfield $\phi$
is the two-component left-handed Weyl spinor
$$
\psi_L=\left(\begin{array}{c}\psi_\a\\0\end{array}\right).
$$
For a Majorana spinor, we have the relation (provided by
$\bar\phi=\phi^\dagger$)
\be
\psi_R=\left(\begin{array}{c}0\\\bar\psi^\ad\end{array}\right)
=(\psi_L)^c = -\gamma^0C{\psi^\dagger_L}^\tau \,. \label{majo}
\ee
}
with Weyl components $\psi_\a$ and $\bar\psi^\ad$.

Consider now the superfield propagator $\langle 0|T\phi(y_1,\theta_1)
\bar\phi(\bar y_2\bar\theta_2)|0\rangle$. Its expansion in powers of
the Grassmann coordinates\footnote{Recall that $y_1$ is a function of
$x_1$, $\theta_1$ and $\bar\theta_1$, and similarly $y_2$ is a function of
$x_2$, $\theta_2$ and $\bar\theta_2$.} includes the Green's functions
\beq
D(x_1 -x_2 )&\equiv& \lvac T z(x_1){\bar z}(x_2)\rvac\, ,
\label{zz1}\\[2mm]
\FF(x_1 -x_2 )&\equiv&\lvac T f(x_1){\bar f}(x_2 )\rvac,
\label{zz2}\\[2mm]
S_{\a}^{\ \bd} (x_1 -x_2 )&\equiv&\lvac T \psi_\a(x_1)
\bar\psi^\bd(x_2 )\rvac,
\label{zz3}
\eeq
as well as Green's functions for the first or second derivatives of
the component fields:
\be
\begin{array}{rcl}
\lvac T \phi(y_1 ,\t_1)\,{\bar\phi}({\bar y}_2 ,\tb_2) \rvac &=&
D(y_1 -{\bar y}_2 )
-2\t_1^\a\tb_{2\,\bd}\,S_{\a}^{\ \bd} (y_1 -{\bar y}_2 )
+\t_1\t_1\,\tb_2\tb_2\,\FF(y_1 -{\bar y}_2 ) \\[2mm]
&=& D(x_1-x_2)
-2\t_1^\a\tb_{2\,\bd}\,S_{\a}^{\ \bd} (x_1-x_2)
+\t_1\t_1\,\tb_2\tb_2\,\FF(x_1-x_2)\label{cac} \\[2mm]
&&+ \,{\rm derivative\,\,terms} .
\end{array}
\ee
A similar expansion can be performed for
$\langle 0|T\phi(y_1,\theta_1)\phi(y_2,\theta_2)|0\rangle$ and for the
conjugate $\langle 0|T\bar\phi(\bar y_1,\bar \theta_1)
\bar\phi(\bar y_2,\bar \theta_2)|0\rangle$.

We now put this system of propagating component fields in a thermal
bath at  finite temperature $T$. The heat bath is expected to affect
propagation. The definitions of the thermal propagators are a straightforward
``thermalization", along the lines of Section \ref{tfbf}, of eqs.
\equ{zz1}-\equ{zz3}:
\beq
D_C(x_1 -x_2 )
&\equiv&
\langle {\sf T}_{\!C} \  z(x_1){\bar z}(x_2)\rangle_\bb
\label{p1}
,\\[2mm]
\FF_C(x_1 -x_2 )
&\equiv&
\langle {\sf T}_{\!C} \  f(x_1){\bar f}(x_2 )\rangle_\bb
\label{p5},\\[1mm]
S_{C\,\a}^{\ \ \ \bd} (x_1 -x_2 )
&\equiv&
\langle {\sf T}_{\!C} \  \psi_\a(x_1)\bar\psi^\bd(x_2 )\rangle_\bb.
\label{p3}
\eeq
Simultaneously, one has to require that each of these thermal propagators for
the components of chiral and antichiral superfields obey KMS conditions. The
relevant condition depends on the statistics of the component fields in the
propagator. That is, thermal propagators of scalar components must obey the
bosonic KMS condition \equ{kmsb},
\beq
D_{C}^>(t_1;\xx_1,t_2;\xx_2) \= \phantom{-}D_{C}^< (t_1+i\b;\xx_1,t_2;\xx_2)
,\label{c1}\\[2mm]
\FF_C ^>(t_1;\xx_1,t_2;\xx_2) \= \phantom{-}\FF_C^< (t_1+i\b;\xx_1,t_2;\xx_2)
,\label{c3}
\eeq
while the thermal propagator of the fermionic components has to satisfy the
fermionic constraint \equ{mmm1},
\be
S_{C\,\a}^{>\ \ \bd} (t_1;\xx_1,t_2;\xx_2)
= -
S_{C\,\a}^{< \ \ \bd} (t_1+i\b;\xx_1,t_2;\xx_2).
\label{c2}
\ee

\subsection{Super-KMS condition and thermal superspace}\label{kmssupf}

In the Introduction, we have motivated the fact that, at finite temperature,
the superspace Grassmann variables should be dependent on imaginary time and
antiperiodic. Consequently, we promote $\t$ and $\tb$ to become time-dependent
coordinates, $\t\rightarrow\th=\th(t)$, $\tb\rightarrow\tbh=\tbh(t)$ with
the antiperiodicity properties \equ{antiper},
\be
{\widehat\t}(t+i\b)=-{\widehat\t}(t),\qquad
{\widehat\tb}(t+i\b)=-{\widehat\tb}(t).
\label{antiper2}
\ee
These conditions induce a {\it temperature-dependent constraint} on the
{\it time-dependent superspace Grassmann coordinates} $\widehat\t(t)$ and
$\widehat\tb(t)$. Notice that while the introduction of a dependence on time
in $\theta$ is a local statement, which should then be visible in the explicit
form of space-time symmetry generators, the above antiperiodicity conditions
are global statements. The latter are not expected to affect symmetry
generators, at least in their classical expressions. But they will appear
in the quantum theory, in the definition of the space of physical states.

We shall call {\it thermal superspace} the space spanned by the variables
$[x^\m,\widehat\t(t),\widehat\tb(t)]$, with $\widehat\t(t)$ and
$\widehat\tb(t)$ obeying the conditions \equ{antiper2}. Note that in taking
superspace Grassmann coordinates $\th(t)$, $\tbh(t)$, we also introduce a
formal time-dependence in the second terms of the variables $y$ and ${\bar y}$
\equ{y}-\equ{ybar}. The implications of this fact will be discussed later on.
For the moment, we only keep track of the $t$-dependence with the notation
$\widehat y_{(t)}$, $\widehat{\bar y}_{(t)}$:
\be
\widehat y^\mu_{(t)}\ \equiv \ x^\m-i \widehat\t(t)\s^\m\widehat\tb(t),
\qquad
\widehat{\bar y}^\mu_{(t)}\ \equiv \ x^\m + i\widehat\t(t)\s^\m\widehat\tb(t).
\label{locy}
\ee

In the present Section, we show that the use of thermal superspace makes it
possible to impose KMS conditions at the level of thermal {\it superfield}
propagators. These superfield KMS conditions will yield the correct bosonic
\equ{c1}--\equ{c3} and fermionic \equ{c2} KMS conditions for the thermal
propagators of the superfield components (paragraph \ref{accompkms}).
Furthermore, we will see how the antiperiodicity \equ{antiper2} allows for a
direct proof of the super-KMS condition in thermal superspace (paragraph
\ref{caproof}).

To start with, we define chiral and antichiral superfields at finite
temperature, denoted by the ``hat" notation $\widehat\phi$, resp.
$\widehat{\bar\phi}$, similarly to \equ{chsexp}, \equ{xexpa}, but with
{\it the thermal superspace Grassmann coordinates $\th(t)$, $\tbh(t)$}
as the expansion parameters. This yields for $\widehat\phi$,
\be
\widehat\phi[\widehat y_{(t)},\widehat\t(t)]=
z[\widehat y_{(t)}]+\sqrt{2}\,\widehat\t(t)\,\psi[\widehat
y_{(t)}] - \widehat\t(t)\widehat\t(t)\, f[\widehat y_{(t)}],
\label{tch}
\ee
whereas for $\widehat{\bar\phi}$ we write
\be
\widehat{\bar\phi}\,[\widehat{\bar y}_{(t)},\widehat\tb(t)]=
{\bar z}[\widehat{\bar y}_{(t)}]+\sqrt{2}\,\,\widehat\tb(t)\,
\bar\psi[\widehat{\bar y}_{(t)}] - \widehat\tb(t)\widehat\tb(t)\,
{\bar f}[\widehat{\bar y}_{(t)}].
\label{tach}
\ee
Since thermal chiral and antichiral superfields are bosonic
objects, and are therefore periodic, these expansions are consistent with the
fact that at finite temperature bosonic fields are periodic in imaginary time,
while fermionic fields are antiperiodic. Moreover, as in the zero-temperature
case, these thermal chiral and antichiral superfields can be seen as solutions
of conditions which generalize $\bar D_\ad \phi =0$, $D_\a{\bar\phi} =0$ [eqs.
\equ{defcs}] to the thermal context. We postpone this point to Section
\ref{tcd}, in which the thermal covariant derivatives are constructed.

Following the same prescription of making $\t$ and $\tb$ time-dependent and
antiperiodic, we next define the chiral-antichiral superfield propagator at
finite temperature to be the thermal generalization of eq.
\equ{cac},
\be
G_C\,[\widehat y_{1(t_1)}, \widehat{\bar y}_{2(t_2)},\widehat\t_1(t_1),
\widehat\tb_2(t_2)] \equiv \langle {\sf T}_{\!C}\  \widehat\phi[\widehat
y_{1(t_1)},\widehat\t_1(t_1)]\,\widehat{\bar\phi}\,[
\widehat{\bar y}_{2(t_2)},\widehat\tb_2(t_2)] \rangle_\bb,
\label{tsdpp}
\ee
and expand it in thermal superspace as
\begin{eqnarray}
G_C[\widehat y_{1(t_1)}, \widehat{\bar
y}_{2(t_2)},\widehat\t_1(t_1),\widehat\tb_2(t_2)]
&=&  D_C[\widehat y_{1(t_1)} - \widehat{\bar y}_{2(t_2)}]
-2\widehat\t_1^\a(t_1)\tb_{2\,\bd}(t_2)\,S_{C\,
\a}^{\ \ \ \bd}[\widehat y_{1(t_1)} - \widehat{\bar y}_{2(t_2)}]
\nonumber \\[2mm]
&&  + \ \widehat\t_1(t_1)\widehat\t_1(t_1)\,
\widehat\tb_2(t_2) \widehat\tb_2(t_2)\,\FF_C [\widehat y_{1(t_1)} -
\widehat{\bar y}_{2(t_2)}]
\nonumber\\[2mm]
&=&  D_C(x_1-x_2)
-2\widehat\t_1^\a(t_1)\tb_{2\,\bd}(t_2)\,S_{C\,
\a}^{\ \ \ \bd}(x_1-x_2)
\nonumber\\[2mm]
&&  + \ \widehat\t_1(t_1)\widehat\t_1(t_1)\,
\widehat\tb_2(t_2) \widehat\tb_2(t_2)\,\FF_C (x_1-x_2) \nonumber\\[2mm]
&& +\,{\rm derivative\,\,terms}.
\label{cact}
\end{eqnarray}
The thermal superfield two-point functions $G_C^>$, resp. $G_C^<$, can be
defined in relation to $G_C$, similarly to \equ{orf}, through
\beq
\hspace{-.6cm}\begin{array}{rll}
G_{C}\,[\widehat y_{1(t_1)},
\widehat{\bar y}_{2(t_2)},\widehat\t_1(t_1),\widehat\tb_2(t_2)] &=&
\t_C(t_1-t_2) \,
G^>_{C}\,[\widehat y_{1(t_1)}, \widehat{\bar y}_{2(t_2)},
\widehat\t_1(t_1),\widehat\tb_2(t_2)]\\[2mm]
&&+\ \t_C(t_2-t_1)\, G^<_{C}\,[\widehat y_{1(t_1)},
\widehat{\bar y}_{2(t_2)},\widehat\t_1(t_1),\widehat\tb_2(t_2)]\,,
\end{array}
\label{orf2}
\eeq
with
\beq
G_C^>\,[\widehat y_{1(t_1)}, \widehat{\bar
y}_{2(t_2)},\widehat\t_1(t_1),\widehat\tb_2(t_2)]\=\langle
\widehat\phi[\widehat y_{1(t_1)},\widehat\t_1(t_1)]\,\widehat{\bar\phi}\,[
\widehat{\bar y}_{2(t_2)},\widehat\tb_2(t_2)] \rangle_\bb
,\label{asup}\\[2mm]
G_C^<\,[\widehat y_{1(t_1)}, \widehat{\bar
y}_{2(t_2)},\widehat\t_1(t_1),\widehat\tb_2(t_2)]\=\langle
\widehat{\bar\phi}
[\widehat{\bar y}_{2(t_2)},\widehat\tb_2(t_2)]
\,\widehat\phi[\widehat y_{1(t_1)}, \widehat\t_1(t_1)]
\rangle_\bb\label{rsup}.
\eeq

We are now equipped to formulate the KMS condition at the level of thermal
superfield propagators. The correct component's KMS conditions
\equ{c1}-\equ{c3} and \equ{c2} can be obtained from the following {\it
superfield KMS (or super-KMS) condition}:
\be
G_C^>\,[\widehat y_{1(t_1)}, \widehat{\bar
y}_{2(t_2)},\widehat\t_1(t_1),\widehat\tb_2(t_2)] =
G_C^<\,[\widehat y_{1(t_1 + i\beta)}, \widehat{\bar
y}_{2(t_2)},\widehat\t_1(t_1+i\b),\widehat\tb_2(t_2)],
\label{skms}
\ee
with the time-translated variable $\widehat y_{1(t_1 + i\beta)}$ given by
\beq\label{y1}
\begin{array}{rcl}
\widehat y_{1(t_1 + i\beta)} &=& \Bigl( t_1+i\beta -i\widehat\t_1(t_1 +
i\b)\s^0\widehat\tb_1(t_1 + i\b)\,\,;\,\,
\xx_1-i\widehat\t_1(t_1 + i\b)\ss\widehat\tb_1(t_1 + i\b)\Bigr) \\[1mm]
&=& \widehat y_{1(t_1)} + (i\beta\,\,;\,\,{\bf 0})\, ,
\end{array}
\eeq
where use has been made of the antiperiodicity conditions \equ{antiper2} for
the thermal superspace coordinates $\widehat\t(t)$ and $\widehat\tb(t)$.

\subsection{Proof of the super-KMS condition}\label{caproof}

Clearly, the superfield KMS condition \equ{skms} is of bosonic type, since
chiral and antichiral superfields are bosonic objects. This condition can be
proven at the superfield level in a way similar to the case of the scalar field
case of Section \ref{tfbf}. At the superfield level, the main ingredients of
the proof are the cyclicity of the thermal trace, the superfield version of the
evolution in imaginary time, and, most essential, the antiperiodicity in
imaginary time of the superspace Grassmann variables.

Let us start by formulating the evolution in imaginary time [eq. \equ{bit}] for
a thermal superfield. Suppose the theory has a hamiltonian $H$ for the fields
$z$, $\psi$ and $f$, which is the generator of time translations for the
$x$-dependence of these fields. Appling this evolution to the various
components in the development of, \eg  a thermal chiral superfield
$\widehat\phi$ [eq. \equ{tch}], we get
\beq
e^{-\b H} \, \widehat\phi[\widehat y_{(t)},\widehat\t(t)] \, e^{\b H}
\=
z[\widehat y^0_{(t)}+i\beta ; \widehat{\vec
y}]+\sqrt{2}\,\widehat\t(t)\,\psi[\widehat y^0_{(t)}+i\beta ; \widehat{\vec
y}]-\widehat\t(t)\widehat\t(t)\,f[\widehat y^0_{(t)}+i\beta ; \widehat{\vec
y}]\nonumber\\[2mm]
\=\widehat\phi[\widehat y^0_{(t)}+i\beta ; \widehat{\vec y},\widehat\t(t)].
\label{sbit2}
\eeq
Similarly, for an antichiral superfield $\widehat{\bar\phi}$, the imaginary
time evolution applied to \equ{tach} yields
\be
e^{-\b H} \, \widehat{\bar\phi}\,[\widehat{\bar y}_{(t)},\widehat\tb(t)] \,
e^{\b H}
=\widehat{\bar\phi}\,[\widehat {\bar y}^0_{(t)}+i\beta ; \widehat{\vec{\bar
y}},\widehat\tb(t)]
\, .
\label{sabit2}
\ee
Because of the antiperiodicity of the Grassmann variables, eq. \equ{antiper2},
and with eq. \equ{y1},
$$
\begin{array}{rcl}
\widehat\phi[\widehat y^0_{(t)}+i\beta ; \widehat{\vec y},\widehat\t(t)]
&=&
\widehat\phi[\widehat y_{(t_1 + i\beta)},\widehat\t(t)],
\\[2mm]
\widehat{\bar\phi}[\widehat {\bar y}^0_{(t)}+i\beta ;
\widehat{\vec {\bar y}},\widehat{\bar\t}(t)]
&=&
\widehat{\bar\phi}[\widehat {\bar y}_{(t_1 + i\beta)},\widehat{\bar \t}(t)].
\end{array}
$$
As a consequence, the thermal superfield evolution is governed by
\beq
e^{-\b H} \, \widehat\phi[\widehat y_{(t)},\widehat\t(t)] \, e^{\b H}
\=
z[ \widehat y_{(t+i\b)}]+\sqrt{2}\,\widehat\t(t)\,\psi[ \widehat y_{(t+i\b)}]
-\widehat\t(t)\widehat\t(t)\, f[ \widehat y_{(t+i\b)}]\nonumber\\[2mm]
\=\widehat\phi[ \widehat y_{(t+i\b)},\widehat\t(t)]
\label{sbit22}
\eeq
for $\widehat\phi$, and by a similar equation for $\widehat{\bar\phi}$:
\be
e^{-\b H} \, \widehat{\bar\phi}[\widehat{\bar y}_{(t)},\widehat\tb(t)] \, e^{\b
H}
=\widehat{\bar\phi}[ \widehat{\bar y}_{(t+i\b)},\widehat\tb(t)]\, .
\label{sbit222}
\ee
Note that, in the above, the time argument of $\widehat\t(t)$ and
$\widehat\tb(t)$ has {\it not} been shifted. The thermal Grassmann
variables -- which are {\it coordinates} -- do not undergo dynamical evolution
in imaginary time generated by the hamiltonian, which only acts on fields,
\ie
\be
e^{-\b H} \, \widehat\t(t) \, e^{\b H}=\widehat\t(t)\, ,\qquad e^{-\b H} \,
\widehat\tb(t) \, e^{\b H}=\widehat\tb(t)
\, .
\label{tghu}
\ee

In order to prove the superfield KMS relation \equ{skms} in thermal superspace,
we start from the thermal superfield two-point function $G_C^>$
\equ{asup},
\be
G_C^>\,[\widehat y_{1(t_1)}, \widehat{\bar
y}_{2(t_2)},\widehat\t_1(t_1),\widehat\tb_2(t_2)] =
{1\over Z(\b)}\, \Tr \Biggl\{ e^{- \beta H}
\widehat\phi
[\widehat y_{1(t_1)},\widehat\t_1(t_1)]\,
\widehat{\bar\phi}\,
[\widehat{\bar y}_{2(t_2)},\widehat\tb_2(t_2)] \Biggr\},\label{conto1}\\[2mm]
\ee
and introduce the thermal component expansions for the superfields [eqs.
\equ{tch}-\equ{tach}]. We then rotate cyclically $\widehat{\bar\phi}$ to the
front, and insert the identity $e^{\beta H}e^{-\beta H}=1$ in the right
side\footnote{To simplify the notation, we occasionally use $\widehat y_i$,
$\th_i$ and $\tbh_i$ instead of $\widehat y_i(t_i)$, $\th_i(t_i)$ and
$\tbh_i(t_i)$
in non ambiguous situations.}:
$$
\begin{array}{l}
\displaystyle{ 1\over Z(\b)}\, \Tr \Biggl\{ e^{- \beta H}
\Bigl(z[\widehat y_1]+\sqrt{2}\,\widehat \t_1\,\psi[\widehat y_1]
-\widehat \t_1\widehat\t_1\, f[\widehat y_1]\Bigr)
\Bigl({\bar z}[\widehat{\bar y}_2]+\sqrt{2}\widehat\tb_2
\bar\psi[\widehat {\bar y}_2]-\widehat \tb_2\widehat\tb_2\,{\bar f}[\widehat
{\bar y}_2]\Bigr) \Biggr\} \nonumber\\[6mm]
\hspace{1.4cm}= \displaystyle{1\over Z(\b)} \Tr\! \Biggl\{
\Bigl({\bar z}[\widehat {\bar y}_2]-\sqrt{2}\widehat \tb_2
\bar\psi[\widehat {\bar y}_2]-\widehat \tb_2\widehat\tb_2\,
{\bar f}[\widehat {\bar y}_2]\Bigr) \\ [2mm]
\hspace{5.2cm}\times
 e^{- \beta H} \Bigl(z[\widehat y_1]+\sqrt{2}\widehat \t_1\psi[\widehat
y_1]-\widehat \t_1
\widehat\t_1\, f[\widehat y_1]\Bigr) e^{\beta H}e^{-\beta H} \Biggr\}.
\end{array}
$$
Here the negative sign in front of one of the fermionic components follows from
the anticommutativity of the Grassmann variables. Indeed, while fermion fields
do not generate a sign when crossed in a cyclic rotation of the quantum
mechanical trace, Grassmann parameters do. We now use time evolution. Upon
cyclically rotating a Boltzmann factor to the front, and using the superfield
evolution eq. \equ{sbit22}, our last expression can be rewritten as:
\be\ba{l}
\!\!\!\displaystyle{1\over Z(\b)}\, \Tr \Biggl\{ e^{- \beta H}
\Bigl({\bar z}[\widehat{\bar y}_2]-\sqrt{2}\,\,\widehat\tb_2\,
\bar\psi[\widehat{\bar y}_2]-\widehat\tb_2\widehat\tb_2\,{\bar
f}[\widehat{\bar y}_2]\Bigr)\nonumber\\[4mm]
\phantom{\displaystyle{1\over Z(\b)}\, \Tr \Biggl\{}
\times\ e^{- \beta H}
\Bigl(z[\widehat y_1] + \sqrt{2}\,\widehat\t_1\,\psi[\widehat
y_1]-\widehat\t_1\widehat\t_1\, f[\widehat y_1]\Bigr) e^{\beta H}
\Biggr\} \nonumber\\[6mm]
=
\displaystyle{1\over Z(\b)}\, \Tr \Biggl\{ e^{- \beta H}
\Bigl({\bar z}[\widehat{\bar y}_{2(t_2)}]-\sqrt{2}\,\,\widehat\tb_2(t_2)\,
\bar\psi[\widehat{\bar y}_{2(t_2)}]-\widehat\tb_2(t_2)\widehat\tb_2(t_2)\,
{\bar f}[\widehat{\bar y}_{2(t_2)}]\Bigr)\nonumber\\[4mm]
\phantom{\displaystyle{1\over Z(\b)}\, \Tr \Biggl\{}
\times \
\Bigl(z[ \widehat y_{1(t_1+i\b)}]  +
\sqrt{2}\,\widehat\t_1(t_1)\,\psi[ \widehat y_{1(t_1+i\b)}]-\widehat\t_1
(t_1)\widehat\t_1(t_1)\,f[ \widehat y_{1(t_1+i\b)}]\Bigr) \Biggr\}\,.
\ea\label{comp2}\ee
At this point, we make use of the antiperiodicity \equ{antiper2}
of the Grassmann variables and set $\widehat\t_1(t_1)=- \widehat \t_1(t_1 +
i\beta)$. As a result, the right side of \equ{comp2} rewrites as
\be\ba{l}
\displaystyle{1\over Z(\b)}\, \Tr \Biggl\{ e^{- \beta H}
\Bigl({\bar z}[\widehat{\bar y}_{2(t_2)}]-\sqrt{2}\,\,\widehat\tb_2(t_2)\,
\bar\psi[\widehat{\bar y}_{2(t_2)}]-\widehat\tb_2(t_2)\widehat\tb_2(t_2)\,
{\bar f}[\widehat{\bar y}_{2(t_2)}]\Bigr)\nonumber\\[4mm]
\hspace{1.1cm}
\times
\Bigl(z[\widehat y_{1(t_1+i\b)}]  -
\sqrt{2}\,\widehat\t_1(t_1+i\b)\,\psi[\widehat
y_{1(t_1+i\b)}]-\widehat\t_1(t_1+i\b)\widehat\t_1(t_1+i\b)\,
f[\widehat y_{1(t_1+i\b)}]\Bigr)\Biggr\}\, .
\ea\label{comp3}\ee
Since fermionic fields do not propagate into bosonic fields, and {\it
vice-versa}, the two negative signs in front of the fermionic components above
can equivalently be replaced by two positive signs. The computation
\equ{conto1}-\equ{comp3} therefore yields
\be\ba{l}
G_C^>\,[\widehat y_{1(t_1)}, \widehat{\bar
y}_{2(t_2)},\widehat\t_1(t_1),\widehat\tb_2(t_2)]
\\[4mm]
=\displaystyle{1\over Z(\b)}\, \Tr \Biggl\{ e^{- \beta H}
\Bigl({\bar z}[\widehat{\bar y}_{2(t_2)}]+\sqrt{2}\,\,\widehat\tb_2(t_2)\,
\bar\psi[\widehat{\bar y}_{2(t_2)}]-\widehat\tb_2(t_2)\widehat\tb_2(t_2)\,
{\bar f}[\widehat{\bar y}_{2(t_2)}]\Bigr)\nonumber\\[4mm]
\times \Bigl(z[\widehat y_{1(t_1+i\b)}]  +
\sqrt{2}\,\widehat\t_1(t_1+i\b)\,\psi[\widehat
y_{1(t_1+i\b)}]-\widehat\t_1(t_1+i\b)\widehat\t_1(t_1+i\b)\,
f[\widehat y_{1(t_1+i\b)}]\Bigr)\Biggr\} \, .
\ea\label{comp33}\ee
Realizing that the second line is just the thermal superfield $\widehat\phi$
\equ{tch} with all time arguments shifted by $i\b$, we rewrite
\equ{comp33} as
\be
G_C^>\,[\widehat y_{1(t_1)}, \widehat{\bar
y}_{2(t_2)},\widehat\t_1(t_1),\widehat\tb_2(t_2)]
=
G_C^<\,[\widehat y_{1(t_1+i\b)}, \widehat{\bar
y}_{2(t_2)},\widehat\t_1(t_1+i\b),\widehat\tb_2(t_2)]\, .
\label{comp4}
\ee
This is the superfield KMS condition \equ{skms}, which is hereby proved.

\subsection{Component KMS conditions from super-KMS}\label{accompkms}

We verify in this paragraph that the superfield KMS condition \equ{skms} yields
the expected component relations \equ{c1}-\equ{c3} and \equ{c2}. As we shall
see, the antiperiodicity of the thermal Grassmann variables is again an
essential ingredient. This is done by expanding in eqs. \equ{asup}--\equ{rsup}
the thermal chiral and antichiral superfields $\widehat\phi$ and
$\widehat{\bar\phi}$ in components, using eqs. \equ{tch} and \equ{tach}:
\beq
G_C^>\,[\widehat y_{1(t_1)}, \widehat{\bar
y}_{2(t_2)},\widehat\t_1(t_1),\widehat\tb_2(t_2)]
\!\!\=\!\!
D_C^>\,[\widehat y_{1(t_1)}, \widehat{\bar y}_{2(t_2)}]
- 2\widehat\t_{1}^\a(t_1)\widehat\tb_{2\, \bd}(t_2) \,S_{C\, \a}^{>\,\
\bd}\,[\widehat y_{1(t_1)},
\widehat{\bar y}_{2(t_2)}] \qquad\nonumber\\[2mm]
&&\!\!\!\!+ \widehat\t_1(t_1)\widehat\t_1(t_1)
\widehat\tb_2(t_2)\widehat\tb_2(t_2)
\,\FF^>_C[\widehat y_{1(t_1)}, \widehat{\bar y}_{2(t_2)}],
\label{aaa}\\[4mm]
G_C^<\,[\widehat y_{1(t_1)}, \widehat{\bar
y}_{2(t_2)},\widehat\t_1(t_1),\widehat\tb_2(t_2)]
\!\!\=\!\!
D_C^<\,[\widehat y_{1(t_1)}, \widehat{\bar y}_{2(t_2)}]
- 2\widehat\t_{1}^\a(t_1)\widehat\tb_{2\, \bd}(t_2) \,S_{C\, \a}^{<\,\
\bd}\,[\widehat y_{1(t_1)},
\widehat{\bar y}_{2(t_2)}] \qquad\nonumber\\[2mm]
&& \!\!\!\!+ \widehat\t_1(t_1)\widehat\t_1(t_1)
\widehat\tb_2(t_2)\widehat\tb_2(t_2)\, \FF^<_C[\widehat
y_{1(t_1)},
\widehat{\bar y}_{2(t_2)}]\, .
\label{rrr}
\eeq
The superfield KMS condition \equ{skms} leads then to the following:

\noindent {\bf (i)} For the scalar component,
\be
D_C^>\,[\widehat y_{1(t_1)}, \widehat{\bar y}_{2(t_2)}] = D_C^<\,[\widehat
y_{1(t_1 + i\b)}, \widehat{\bar y}_{2(t_2)}],
\label{c111}
\ee
which reduces, when returning to variables $x=(t;\xx)$, to the bosonic KMS
relation \equ{c1}, $D_{C}^>(t_1;\xx_1,t_2;\xx_2) = D_{C}^<
(t_1+i\b;\xx_1,t_2;\xx_2)$.

\noindent {\bf (ii)} For the fermionic component,
\be
\widehat\t_1^\a(t_1)\widehat\tb_{2\, \bd}(t_2) \,S_{C\, \a}^{>\,\
\bd}\,[\widehat y_{1(t_1)},
\widehat{\bar y}_{2(t_2)}] =
\widehat\t_1^\a(t_1+i\beta)\widehat\tb_{2\, {\bd}}(t_2) \,S_{C\, \a}^{<\,\
\bd}\,[\widehat y_{1(t_1 + i\b)} , \widehat{\bar y}_{2(t_2)}].
\label{fff}
\ee
With the antiperiodicity condition \equ{antiper2},
$\widehat\t_1^\a(t_1+i\beta)=-\widehat\t_1^\a(t_1)$, we obtain
\be
S_{C\, \a}^{>\,\ \bd}\,[\widehat y_{1(t_1)}, \widehat{\bar y}_{2(t_2)}]
=
- S_{C\, \a}^{<\,\ \bd}\,[\widehat y_{1(t_1 + i\b)}, \widehat{\bar
y}_{2(t_2)}],
\label{crs}
\ee
which yields, in the variables $x=(t;\xx)$, the fermionic KMS condition
\equ{c2} with the correct relative sign,
$S_{C\, \a}^{>\,\ \bd}\,(t_1;\xx_1,t_2;\xx_2)=
- S_{C\, \a}^{<\,\ \bd}\,(t_1+i\b;\xx_1,t_2;\xx_2)$.

\noindent {\bf (iii)} For the auxiliary field, one gets from \equ{skms}:
\be
\ba{rcl}
&&\widehat\t_1(t_1)\widehat\t_1(t_1)
\widehat\tb_2(t_2)\widehat\tb_2(t_2)
\,\FF^>_C[\widehat y_{1(t_1)},
\widehat{\bar y}_{2(t_2)}] \\[2mm]
&&\hspace{2cm}=
\widehat\t_1(t_1+i\b)\widehat\t_1(t_1+i\b)
\widehat\tb_2(t_2)\widehat\tb_2(t_2)\, \FF^<_C[\widehat
y_{1(t_1 + i\b)},
\widehat{\bar y}_{2(t_2)}].\qquad
\ea
\label{gthf}
\ee
With $\widehat\t_1(t_1 + i\b)= -\widehat\t_1(t_1)$ \equ{antiper2} and
in the variables $x=(t;\xx)$, this is the bosonic KMS condition \equ{c3},
$\FF^>_C(t_1;\xx_1,t_2;\xx_2)=\FF^<_C(t_1+i\b;\xx_1,t_2;\xx_2)$.

Finally, we recall that we have considered here only the chiral-antichiral
thermal superfield two-point function which, at zero temperature, contains the
kinetic propagators for the scalar and spinor fields. Mass contributions to
propagators would arise from the chiral-chiral Green's function $\langle T_C
\widehat\phi\,\widehat\phi\rangle_\bb$, or from its conjugate $\langle T_C
{\widehat{\bar\phi}}\,{\widehat{\bar\phi}}\rangle_\bb$. To each of these cases,
there corresponds a superfield KMS condition. Since their treatment on thermal
superspace is entirely similar to the chiral-antichiral case discussed above,
we don't consider them explicitly here.

\section{Thermal covariant derivatives and thermal\hfill\break
supercharges}\label{tcd}

Up to this point, our construction of thermal superspace has been somewhat
academic. We have simply translated in a superfield formalism results of finite
temperature field theory for the fields of the chiral supermultiplet. The price
has been to introduce a dependence on time/temperature in Grassmann superspace
coordinates, and also in chiral space-time coordinates $y^\mu$ and $\bar
y^\mu$. In contrast to the antiperiodic $\widehat\theta$ variables, the latter
dependence is periodic in $t\rightarrow t+i\beta$ since $y^\mu$ and $\bar
y^\mu$ are bosonic quantities. The main interest in superspace lies however in
the natural representation it provides for the super-Poincar\'e algebra in
terms of derivatives with respect to superspace coordinates. A function on
superspace -- a superfield -- carries then automatically a representation of
supersymmetry. The purpose of this section is to construct the supersymmetry
generators and covariant derivatives on thermal superspace. Their existence is
not a surprise. The algebra reflects local properties in superspace while the
distinction between fermions and bosons at finite temperature is related to a
global property: periodicity or antiperiodicity of fields around the
temperature circle with radius $\beta/2\pi$. However, {\it the existence of a
supersymmetry algebra on thermal superspace should not be assimilated to a
statement that supersymmetry does not break at finite $T$}. That such an
algebra exists does  {\it not} imply that a supersymmetric field theory can be
constructed carrying the same symmetry algebra. We shall come back to this
point in Section \ref{tbbss}.

Deriving expressions for the supercharges and the covariant derivatives on
thermal superspace can be done simply by performing the change of variables
from usual, zero temperature, superspace to thermal superspace, \ie:
$$
(x^\m,\t,\tb)\quad \longrightarrow\quad
\left(x'^\m=x^\m,\t'=\th(t),\tb'=\tbh(t)\right)\, ,
$$
with $t=x^0$. Under this change of variables, the partial derivatives with
respect to $\xx$, $\t$ and $\tb$ transform trivially,
\beq
\pad{}{\xx} &\longrightarrow& \pad{}{\xx'} = \pad{}{\xx}\, ,\label{ttxx}\\[2mm]
\pad{}{\t^\a}&\longrightarrow&\pad{}{\t'^\a}=\pad{}{\th^\a}
,\label{ttta}\\[2mm]
\pad{}{\tb^\ad}&\longrightarrow&\pad{}{\tb'^\ad}=\pad{}{\tbh^\ad}
,\label{tttba}
\eeq
while the time derivative has to take the time-dependence of the thermal
Grassmann variables into account:
\be
\pad{}{t} \longrightarrow \pad{}{t'} + \pad{\th^\a}{t}\pad{}{\th^\a} +
\pad{\tbh^\ad}{t}\pad{}{\tbh^\ad}\,, \qquad
\left(\pad{t'}{t}=1\right)\, .\label{fintt}
\ee
Consequently, we define the partial time derivative  at finite temperature as
\be
\widehat{\pad{}{t}} \ \equiv \ \pad{}{t} - \Delta\, . \label{finttt}
\ee
We call this object the {\it thermal time derivative};  $\Delta$ accounts for
the thermal corrections,
\be
\Delta = \pad{\th^\a}{t}\pad{}{\th^\a} + \pad{\tbh^\ad}{t}\pad{}{\tbh^\ad}.
\label{DD}
\ee
Accordingly, we also define a {\it thermal space-time derivative} as
\be
\widehat\pa_\m=
\left( \pad{}{t} - \Delta \ \ ; \ \pad{}{\xx}\right)\,.
\label{thstd}
\ee

Let us now construct the thermal covariant derivatives. We proceed by
replacing, in the expressions of the (zero-temperature) covariant derivatives
\equ{covder}, the zero-temperature Grassmann variables and derivative operators
by their thermal counterparts. This means that (i) we replace the
zero-temperature, constant Grassmann parameters $\t$, $\tb$ by the thermal,
time-dependent and antiperiodic parameters $\th$, $\tbh$, and that (ii) the
derivative operators $\pa_\m$, $\pa/\pa{\t}$ and $\pa/\pa{\tb}$ are replaced
by $\widehat\pa^\m$, $\pa/\pa{\th}$ and $\pa/\pa{\tbh}$. This yields {\it
thermal covariant derivatives} $\widehat D_\a$ and $\widehat{\bar D}_\ad$ in
the form:
\be
\widehat D_\a = \pad{}{\th^\a} -i\,\s^\m_\aad\tbh^\ad \widehat\pa_\m,
\qquad \qquad
\widehat{\bar D}_\ad = \pad{}{\tbh^\ad} -i\, \th^\a\s^\m_\aad \widehat\pa_\m
,\label{thcdc}
\ee
which write explicitly, using eqs. \equ{thstd} and \equ{DD}, as
\beq
\widehat D_\a \= \pad{}{\th^\a}
-i\,\s^\m_\aad\tbh^\ad \pa_\m +i\,\s^0_\aad\tbh^\ad\left(
\pad{\th^\g}{t}\pad{}{\th^\g} + \pad{\tbh^\gd}{t}\pad{}{\tbh^\gd} \right)
,\label{thcd}\\[2mm]
\widehat{\bar D}_\ad \= \pad{}{\tbh^\ad}
-i\, \th^\a\s^\m_\aad \pa_\m +i\,\s^0_\aad\tbh^\ad\left(
\pad{\th^\g}{t}\pad{}{\th^\g} + \pad{\tbh^\gd}{t}\pad{}{\tbh^\gd} \right)
.\label{thcdb}
\eeq
In order to validate these expressions, we observe that they play, in thermal
superspace, the same role as the usual covariant derivatives of supersymmetry
in $T=0$ superspace.

Firstly, the thermal covariant derivatives obey the same anticommutation
relations as at $T=0$. This can readily be checked by direct computation of the
anticommutators. One obtains, in perfect analogy to the $T=0$ case,
\be
\{ \widehat D_\alpha , \widehat{\bar D}_\ad \} =
-2i\sigma^\mu_{\alpha\ad}\widehat\partial_\mu,\qquad\qquad
\{\widehat D_\a\,,\, \widehat D_\b \} = \{ \widehat{\bar D}_\ad \,,\,
\widehat{\bar D}_\bd \} =0.\label{accd}
\ee
This is actually obvious upon noticing that the thermal space-time derivative
$\widehat\pa_\m$ gives zero when acting on the $t$-dependent variables $\th$,
$\tbh$, since
\be
\widehat\pa_0\,\th^\a= \pad{\th^\a}{t} - \pad{\th^\g}{t}\delta^\a_\g =0
,\qquad\qquad
\widehat\pa_0\,\tbh^\ad= \pad{\tbh^\ad}{t} -
\pad{\tbh^\gd}{t}\delta^\ad_\gd =0,
\label{obs1}
\ee
and plays therefore the same role for the thermal Grassmann variables as that
of the usual space-time derivative for the $t$-independent, non thermal $\t$
and $\tb$. In this sense, the thermal time (and consequently the thermal
space-time) derivative is a covariantization, with respect to thermal
superspace, of the zero-temperature time (space-time) derivative.

Secondly, the thermal covariant derivatives $\widehat{\bar D}_\ad$ and
$\widehat D_\a$ provide a {\it definition} of the thermal chiral and antichiral
superfields $\widehat\phi$ and $\widehat{\bar\phi}$, eqs. \equ{tch} and
\equ{tach}, as the solution to the thermal generalization of the conditions
\equ{defcs}:
\be
\widehat{\bar D}_\ad\ \widehat\phi = 0,\qquad \qquad
\widehat D_\a\ \widehat{\bar\phi}= 0\, .
\label{defpbxt}
\ee
We could have actually derived the thermal covariant derivatives directly from
the requirements
$$
\widehat{\bar D}_\ad\ \widehat y^\mu = 0,\qquad \qquad
\widehat D_\a\ \widehat{\bar y}^\mu= 0,
$$
which are equivalent to the chirality conditions \equ{defpbxt}.

Quite naturally, our next aim is to derive the thermal supersymmetry charges.
The zero-temperature supercharges are, in our conventions:
\be
Q_\a = -i\,\pad{}{\t^\a} + \s^\m_\aad\tb^\ad \pa_\m\, ,\qquad\qquad
\bar Q_\ad =\phantom{-} i\,\pad{}{\tb^\ad} - \t^\a\s^\m_\aad \pa_\m.
\label{zthsb}
\ee
The corresponding thermal objects are constructed using the same procedure as
the one used above for the thermal covariant derivatives. This yields the
following expressions for the {\it thermal supercharges}:
\beq
\qh_\a \= -i\,\pad{}{\th^\a} + \s^\m_\aad\tbh^\ad \pa_\m -
\s^0_\aad\tbh^\ad\left( \pad{\th^\g}{t}\pad{}{\th^\g} +
\pad{\tbh^\gd}{t}\pad{}{\tbh^\gd} \right)\, ,\label{ths}\\[2mm]
\qbh_\ad \=\phantom{-} i\,\pad{}{\tbh^\ad} - \th^\a\s^\m_\aad \pa_\m +
\s^0_\aad\tbh^\ad\left( \pad{\th^\g}{t}\pad{}{\th^\g} +
\pad{\tbh^\gd}{t}\pad{}{\tbh^\gd} \right)\, ,\label{thsb}
\eeq
or, in a compact form,
\be
\qh_\a = -i\,\pad{}{\th^\a} + \s^\m_\aad\tbh^\ad \widehat\pa_\m
\, ,\qquad\qquad
\qbh_\ad = i\,\pad{}{\tbh^\ad} - \th^\a\s^\m_\aad \widehat\pa_\m
\, .\label{thsc}
\ee
It is again straightforward to check that thermal supercharges obey the same
anticommutation relations with thermal covariant derivatives as at $T=0$:
\be
\{ \qh_\a \,,\, \dh_\b \} =
\{ \qbh_\ad \,,\, \dh_\b \} =
\{ \qh_\a \,,\, \dbh_\bd \} =
\{ \qbh_\ad \,,\, \dbh_\bd \} = 0\, .\label{qdqd}
\ee
Furthermore we have
\be
\{ \widehat Q_\alpha , \widehat{\bar Q}_\ad \} =
2i\sigma^\mu_{\alpha\ad}\widehat\partial_\mu,\qquad\qquad
\{ \qh_\a \,,\, \qh_\b \} = \{ \qbh_\ad \,,\, \qbh_\bd \}=0\, .\label{qqqq}
\ee

\section{The thermal super-Poincar\'e algebra}\label{tssa}

In order to compute the full thermal super-Poincar\'e algebra, we need, in
addition to the thermal supercharges constructed in the previous section,
expressions for the thermal translations and thermal Lorentz generators. Let us
start by writing the expressions in the zero-temperature case, \ie for the
generators of the $T=0$ supersymmetry algebra acting on $T=0$ superfields. The
$T=0$ translation generators act on a chiral superfield $\phi$ simply as:
\beq
[ P^\m,\phi(x,\t,\tb)]=-i\,\pa^\m\phi(x,\t,\tb),
\label{pss1}
\eeq
while the Lorentz generators entail a $\t$-, $\tb$-dependent part:
\beq
\begin{array}{rcl}
[M^\mn,\phi(x,\t,\tb)] &=&
\Bigl[i\,(x^\m\pa^\n -x^\n\pa^\m) \\[1mm]
&&\hspace{1.3cm}
+ \displaystyle{i\over 2}\,
(\s^\mn)_\a^{\ \b}\t_\b\pad{}{\t_\a}
+ \displaystyle{i\over 2}\,
(\bar\s^{\n\m})^\ad_{\ \bd}\tb^\bd\pad{}{\tb^\ad}\Bigr]\phi(x,\t,\tb).
\label{mss1}
\end{array}
\eeq
Notice that the above equations display Poincar\'e transformations of the
(Lorentz) scalar superfield at a fixed superspace coordinate point
$(x,\theta,\bar\theta)$. In other words, they display
$\phi'(x,\theta,\bar\theta)-\phi(x,\theta,\bar\theta)$ for a Poincar\'e
transformation
$$
(x,\theta,\bar\theta) \,\,\longrightarrow\,\,(x',\theta',\bar\theta')\,,
\qquad
\phi(x,\theta,\bar\theta) \,\,\longrightarrow\,\,
\phi'(x',\theta',\bar\theta')\,.
$$

At finite temperature, the translation and Lorentz generators above are to be
modified -- similarly to the thermal covariant derivatives and the thermal
supercharges -- by replacing $\pa_\m$ with $\widehat\pa_\m$, and $\t$, $\tb$
with $\widehat\t(t)$, $\widehat\tb(t)$. Therefore we define the action of {\it
thermal translation} and {\it thermal Lorentz generators} on a thermal scalar
superfield through
\beq
[ \widehat P^\m , \widehat \phi(x,\th,\tbh)] =
-i\,\widehat \pa^\m\widehat \phi(x,\th,\tbh),\label{ppss1}
\eeq
and
\beq
\begin{array}{rcl}
[\widehat M^\mn,\widehat \phi(x,\th,\tbh)] &=&
\Biggl[i(x^\m\widehat \pa^\n -x^\n\widehat \pa^\m) \\[1mm]
&&\hspace{1.3cm}
+ \displaystyle{i\over 2}
(\s^\mn)_\a^{\ \b}\th_\b\pad{}{\th_\a}
+ \displaystyle{i\over 2}\,
(\bar\s^{\n\m})^\ad_{\ \bd}\tbh^\bd\pad{}{\tbh^\ad}\Biggr]
\widehat\phi(x,\th,\tbh).
\end{array}
\label{mmss1}
\eeq

As only the derivative in the time direction is modified at finite temperature,
we now distinguish between the generators which are genuinely thermal [that is,
which involve the operator $\Delta$ introduced in eq. \equ{DD}] and those
generators of which the only thermal character comes from the superspace
coordinates being time-dependent. We drop the ``hat" for the latter operators,
and hence decompose the thermal translations $\widehat P^\m$ into $\widehat
P^0$ and $P^i$, while the thermal Lorentz generators $\widehat M^{\mn}$ are
separated into thermal Lorentz boosts $\widehat M^{0i}$ and rotations $M^{ij}$.
A straightforward computation of the commutation rules yields the thermal
Poincar\'e algebra -- the bosonic sector of the thermal super-Poincar\'e
algebra:
\beq
[\widehat M^{0i},\widehat P^0]\=-i\, P^i\, ,\label{pp1}\\[2mm]
[\widehat M^{0i}, P^j]\= \phantom{-} i\, \eta^{ij} \widehat P^0\, ,
\label{pp2}\\[2mm]
[M^{ij},\widehat P^0]\=\phantom{-}0\, ,\label{pp3}\\[2mm]
[M^{ij},P^k]\=-i\,(\eta^{ik} P^j - \eta^{jk} P^i)\, ,\label{pp4}\\[2mm]
[M^{ij}, M^{kl}]\= -i\,( \eta^{ik}M^{jl} + \eta^{jl}M^{ik} -
\eta^{il}M^{jk} - \eta^{jk}M^{il} )
\, ,\label{pp5}\\[2mm]
[\widehat M^{0i}, M^{jk}]\= -i\,( \eta^{ik}\widehat M^{0j} -
\eta^{ij}\widehat M^{0k} )
\, ,\label{pp6}\\[2mm]
[\widehat M^{0i}, \widehat M^{0j}]\= -i\, M^{ij}\, ,\label{pp7}\\[2mm]
[\widehat P^0, \widehat P^0] \=
[\widehat P^0, P^i] \ =\
[ P^i, P^j] \ = \ 0\, . \label{pp8}
\eeq
The fermionic sector is given by
\beq
\{ \qh_\a , \qbh_\bd\} \= -2\, \left(\s^0_{\a\bd} \widehat P_0
- \s^i_{\a\bd}P_i \right),\label{pp9}\\[2mm]
[\widehat P^0,\qh_\a]\= [\widehat P^0;\qbh_\ad] \ =\ [P^i;\qh_\a]
\ =\ [P^i,\qbh_\ad]\ =\ 0,\label{pp10}\\[2mm]
[\widehat M^{0i},\qh_\a] \= -{i\over 2}(\s^{0i})_\a^{\ \b}\, \qh_\b
,\label{pp11}\\[2mm]
[\widehat M^{0i},\qbh_\ad] \= -{i\over 2}(\bar\s^{0i})_{\ad\bd}\, \qbh^\bd
,\label{pp12}\\[2mm]
[M^{ij},\qh_\a] \= -{i\over 2}(\s^{ij})_\a^{\ \b}\, \qh_\b
,\label{pp13}\\[2mm]
[M^{ij},\qbh_\ad] \= -{i\over 2}(\bar\s^{ij})_{\ad\bd}\, \qbh^\bd.
\label{pp14}
\eeq
The {\it thermal super-Poincar\'e algebra} \equ{pp1}---\equ{pp14} has hence the
same structure as at $T=0$, and contains the same number of supercharges. This
is a natural result of our construction which introduces a local dependence on
time/temperature on superspace coordinates. The algebra is maintained by the
appropriate covariantizations.

An interpretation of the role of the thermal time translation operator
$\widehat P^0=-i\widehat\pa^0$ -- the thermal covariantization of $P^0$ -- can
be given as follows. Clearly, in the thermal Poincar\'e algebra
\equ{pp1}-\equ{pp8}, the thermal Lorentz boosts $\widehat M^{0i}$ mix space-
and time-translations. Even if a {\it covariant} formulation can be given
\cite{weldon}, immersing the field theory in a heat bath via periodicity or
antiperiodicity conditions on (imaginary) time removes {\it invariance} under
the Lorentz boosts \cite{aoyama,ojima}. Without these boosts, the Poincar\'e
algebra reduces to spatial rotations $M^{ij}$ and translations $\widehat P^\m$,
which are true symmetries of finite temperature field theory. The thermal
supersymmetry generators $\widehat Q_\alpha$ and $\widehat{\bar Q}_\ad$ then
add to these space-time symmetries to form a three-dimensional supersymmetry
algebra in which the thermal time translation operator $\widehat P^0$ is a
central charge: it commutes with all operators $P^i$, $M^{ij}$, $\widehat
Q_\alpha$ and $\widehat{\bar Q}_\ad$.

\section{Thermal supersymmetry transformations\hfill\break of
bosons and fermions}\label{ttsstt}

We wish here to compute the transformations of thermal superfield components
under the thermal supersymmetry transformations generated by the thermal
supercharges $\qh_\ad$ and $\qbh_\ad$, eqs. \equ{ths}-\equ{thsb}. This means
translating into component language the thermal superfield transformation
$\widehat\delta$ given by
\be
\widehat\delta\widehat\phi(x,\th,\tbh)= i
\left(
\widehat\epsilon^\a\qh_\a +\widehat{\bar\epsilon}_\ad\qbh^\ad
\right)
\widehat\phi(x,\th,\tbh).
\label{tsstrsf}
\ee
Recall that at zero temperature, this is most easily done for a chiral
superfield $\phi$ in chiral variables $y$, $\t$. Since
\be
Q_\a y^\m=\qb_\ad\overline y^\m=0\, ,
\qquad Q_\a \overline y^\m = 2(\s^\m\tb)_\a\,,
\qquad \qb_\ad y^\m = -2(\t\s^\m)_\ad\,,
\label{t0rel}
\ee
one easily deduces that the $T=0$ supercharges acting on $\phi(y,\theta)$ are
\be
Q_\a\phi(y,\t)=-i\pad{}{\t^\a}\phi(y,\t)\, ,\qquad
\qb_\ad\phi(y,\t)=-2(\t\s^\m)_\ad\pad{}{y^\m}\phi(y,\t).
\label{ytt0}
\ee
It is then immediate to compute the supersymmetry transformations of the
components of $\phi$ in  chiral variables by expanding $\delta\phi(y,\t)= i
(\epsilon^\a Q_\a +\bar\epsilon_\ad\qb^\ad)\phi(y,\t)$.

The aim here is to investigate the thermal situation. By construction, thermal
supercharges in the thermal chiral variables $\widehat y$, $\th$, $\tbh$ are
analogous to the $T=0$ expressions above. We also have
\be
\qh_\a \widehat\t^\b=-i\delta^\b_\a,
\qquad\qbh_\ad \tbh^\bd=i\delta^\bd_\ad,
\qquad \qh_\a \tbh^\bd = 0,
\qquad \qbh_\ad \widehat\t^\b = 0,
\label{obstt}
\ee
and
\be
\widehat\pa_\n\yh^\m=\delta_\n^\m,\qquad
\widehat\pa_\n\widehat{\overline y}^\m=\delta_\n^\m.
\label{qyhg}
\ee
As a consequence, one sees that
\be
\qh_\a \widehat y^\m=\qbh_\ad\widehat{\overline y}^\m=0,
\qquad \qh_\a \widehat{\overline y}^\m = 2(\s^\m\tbh)_\a,
\qquad \qbh_\ad \widehat y^\m = -2(\th\s^\m)_\ad.
\label{tfrel}
\ee
Therefore, the supercharges $\qh_\a$ and $\qbh_\ad$, when acting on a
general\footnote{A function of the thermal superspace coordinates  $\widehat
y$, $\th$ and $\tbh$.} thermal superfield denoted $\widehat F$ expressed in the
chiral variables, yield
\beq
\qh_\a\widehat{F}(\widehat y, \th,\tbh) \= -i
\pad{}{\th^\a}\widehat{F}(\widehat y, \th,\tbh),\label{wh1}\\[2mm]
\qbh_\ad\widehat{F}(\widehat y, \th,\tbh) \= -i \left(
-\pad{}{\tbh^\ad}-2i(\th\s^\m)_\ad\pad{}{\yh^\m}\right) \widehat{F}
(\widehat y, \th,\tbh).\label{wh2}
\eeq
On a thermal chiral superfield $\widehat\phi(\yh,\th)$, these thermal
supercharges obviously reduce to
\be
\qh_\a\widehat\phi(\yh,\th) = -i \pad{}{\th^\a}\widehat\phi(\yh,\th),\qquad
\qbh_\ad\widehat\phi(\yh,\th) = -2(\th\s^\m)_\ad\pad{}{\yh^\m}
\widehat\phi(\yh,\th),\label{wch}
\ee
which is nothing but the thermal version of eqs. \equ{ytt0}. Analogous
expressions can be constructed for the thermal supercharges acting on
antichiral thermal superfields as functions of variables $\widehat{\bar y}$ and
$\tbh$.

Inserting the supercharges \equ{wch} into the thermal supersymmetry
transformation \equ{tsstrsf} leads to
\be
\widehat\delta\widehat\phi(\yh,\th)=
\left(
\widehat\epsilon^\a\pad{}{\th^\a}
-2i(\th\s^\m\widehat{\bar\epsilon})\pad{}{\yh^\m}
\right)
\widehat\phi(\yh,\th)\,.
\label{ttrf}
\ee
Defining then $\pad{}{\yh^\m}
\varphi(\yh) \equiv \pa_\m \varphi$, for $\varphi=z$ or $\psi$, we get:
\be
\widehat\delta\widehat\phi(\yh,\th)
= \widehat\epsilon^\a\left[\sqrt{2}\psi_\a(\yh)
-2\th_\a f(\yh)\right] -2i(\th\s^\m\widehat{\bar\epsilon})
\left[\pa_\m z(\yh) +\sqrt{2}\th^\a\pa_\m\psi_\a(\yh)\right].
\label{ttrfd}
\ee
Comparison with the component expansion of
$\widehat\delta\widehat\phi(\yh,\th)$ immediately leads to:
\beq
\widehat\delta z\=
\sqrt{2}\widehat\epsilon^\a \psi_\a \,,
\label{tt1}\\[2mm]
\widehat\delta \psi_\a\=
-\sqrt{2}\widehat\epsilon_\a f -i\sqrt{2}
(\s^\m \widehat{\bar\epsilon})_\a (\pa_\m z)\, ,
\label{tt2}\\[2mm]
\widehat\delta f
\=-i\sqrt{2} (\s^\m \widehat{\bar\epsilon})^\a (\pa_\m \psi_\a)\, .
\label{tt3}
\eeq
In these transformations of the chiral multiplet, the unique difference with
the case of zero temperature is the appearance of the thermal spinorial
parameter $\widehat\epsilon$ in place of the constant spinorial parameter
$\epsilon$ of $T=0$ supersymmetry.

The nature of $\widehat\epsilon$  is however deeply related to finite
temperature. Reversing the argument given in the Introduction, it is
clear from  \equ{wch} that the supercharge $\widehat Q_\alpha$ acting on
$\widehat\phi$ translates $\widehat\theta$ by an amount $\widehat\epsilon$.
Since however the Grassmann coordinate $\widehat\theta$ is time-dependent and
antiperiodic, one must then assume that $\widehat\epsilon$ itself is a
time-dependent, antiperiodic spinor:
\beq
\label{epst}
\widehat\epsilon\,(t+i\beta) = -\widehat\epsilon\,(t)\,.
\eeq
This is confirmed by the action on $\widehat\phi$ of the conjugate supercharge
$\widehat{\bar Q}_\ad$ [eq. \equ{wch}], which translates $\widehat y^\mu$ by an
amount $-2i(\widehat\theta\sigma^\mu\widehat{\bar\epsilon})$. In order that the
time-dependence of $\widehat y$ remains periodic, $\widehat\epsilon$ must be
antiperiodic, as in eq. \equ{epst}.

The time-dependence of $\widehat\epsilon$ is, in this thermal superspace
formalism, the manifestation of the breaking of global supersymmetry at finite
temperature. Computing the commutator $[\widehat\delta_1,\widehat\delta_2]$ of
two thermal supersymmetry transformations $\widehat\delta_1$,
$\widehat\delta_2$ will not close the algebra we have derived simply because
$$
\{\widehat Q_\alpha, \widehat\epsilon(t)\} \ne 0 \ne
\{ \widehat{\bar Q}_\ad,\widehat\epsilon(t)\} \,,
$$
in contrast with the zero-temperature case in which $\epsilon$ is a constant
spinor. Notice also that the supersymmetry transformation of the highest
component $f$ of the superfield $\widehat\phi$ is not a space-time derivative
since $\partial_\mu\widehat\epsilon\ne0$. The method usually applied to
construct invariant lagrangians by tensor calculus will no longer hold at
finite temperature. And actions invariant under supersymmetry at zero
temperature will exhibit a breaking pattern which can easily be studied using
thermal superspace.

\section{Thermal supersymmetry and the \hfill\break Wess-Zumino
model}\label{tbbss}

%\subsection{Thermal reduction and the action of the Matsubara
%modes}\label{tsbofe}

In looking for  realizations of thermal supersymmetry, we shall be
applying the transformations generated by the thermal charges to systems of
thermal fields. And thermal fields are known to be characterized by {\it
global} periodicity conditions which distinguish bosons from fermions. Indeed,
a thermal bosonic field (say $z$) is periodic in imaginary time, while a
thermal fermionic field (denoted $\psi$) is antiperiodic~:
\be
z(t+i\b,\xx)=z(t,\xx)\,,\qquad\qquad\psi(t+i\b,\xx)=-\psi(t,\xx)\,.
\label{perb}
\ee
These thermal characterizations are actually equivalent to the bosonic, resp.
fermionic KMS conditions \equ{kmsb}, \equ{kmsf}, which are hence of global
nature as well.  Therefore, we expect to see signs of supersymmetry breaking
when realizing the thermal supersymmetry algebra, which is a local structure,
on periodic (bosonic), and
antiperiodic (fermionic) fields.  A common way
of introducing the fields' global periodicity properties is to develop them
thermally {\it \`a la} Matsubara. In the Matsubara expansion, bosons are
expanded in thermal modes as
\be
z(t,\xx)={1\over\sqrt{\b}}\sum_{n=-\infty}^{n=\infty} z_n(\xx) \ e^{i\o_n^B t}\
,\label{bd}
\ee
where
\be
\o_n^B= {2n\pi\over \b}\label{bosfreq}
\ee
are the bosonic Matsubara frequencies, while fermions are developed as
\be
\psi(t,\xx)={1\over\sqrt{\b}}\sum_{n=-\infty}^{n=\infty} \psi_n(\xx) \
e^{i\o_n^F t}\,,\label{fd}
\ee
with the fermionic Matsubara frequencies
\be
\o_n^F= {(2n+1)\pi\over\b}\,.\label{fermfreq}
\ee
Clearly, these developements contain the information on the periodicity in
time.

The Matsubara expansion, after rotation to euclidean time, is a realization
of the imaginary time formalism for finite temperature field theory. It is
an expansion on $S^1\times \real^3$, the circle $S^1$ having length $\beta
= 1/T$. In a supergravity theory, it could be regarded as
a particular Scherk-Schwarz compactification \cite{SchSch} scheme of the
time direction.

Since we have considered only non-interacting scalar and fermionic matter
fields described by chiral and antichiral superfields, the
natural zero-temperature limiting lagrangian density -- to be studied now
at finite temperature -- is that of the free (off-shell) Wess-Zumino
model
\be
S^{d=4} = \intq \bigl( \LL^{d=4}_{\rm kin} + \LL^{d=4}_{ \rm mass}\bigr),
\label{acto}
\ee
with kinetic and mass lagrangians given by,
\beq
\LL^{d=4}_{\rm kin} \=  \half(\pa_\m A)^2 + \half(\pa_\m B)^2 + {i\over 2}
\bar\psi \g^\m\pa_\m \psi + \half (F^2+G^2)\,,\label{kinl}\\[2mm]
\LL^{d=4}_{ \rm mass} \=  -M_4(\half \bar\psi \psi + AF + BG)\,.\label{lm}
\eeq
$M_4$ is the mass, $\psi$ a four-component Majorana fermion, $A$, $B$, $F$ and
$G$ are real scalars. The auxiliary fields $F$ and $G$ obey the equations of
motion
\be
F=M_4\,A\,,\qquad\qquad G=M_4\, B\,.
\label{eqmto}
\ee
One can equivalently use complex scalar fields $z$ and $f$ with
\be
z(x)={1\over\sqrt{2}}[A(x)+i B(x)]\,,\qquad\qquad
f(x)={1\over\sqrt{2}}[F(x)+iG(x)]\,.
\label{abz}
\ee
The supersymmetry transformations at $T=0$ are given by:
\be\ba{rclcrcl}
\d A \= \bar\epsilon \psi\,,&& \d B \= i\bar\epsilon\g_5\psi\,,\\[3mm]
\d F \= i\bar\epsilon\g^\m\pa_\m \psi\,,&& \d G \=  -\bar\epsilon\g_5
\g^\m\pa_\m \psi\,,\\[3mm]
\d\psi \= -[i\g^\m\pa_\m(A+iB\g_5)+F+iG\g_5]\epsilon\,, &&\\[2mm]
\d\bar\psi \= -\bar\epsilon [i\g^\m\pa_\m(-A+iB\g_5)+F+iG\g_5]\,.&&
\ea\label{ss2}\ee
Recall that under these $T=0$ transformations, the kinetic and mass
contributions to the action $S^{d=4}$ are separately invariant, \ie
$\d\int d^4x\,\LL^{d=4}_{\rm kin} = \d\int d^4x\,\LL^{d=4}_{ \rm mass} = 0$.
Concretely, omitting in each case a space-time derivative
which integrates to zero,
\beq
\label{transf1}
\begin{array}{rcl}
\displaystyle{\d\int d^4x\,\LL^{d=4}_{\rm kin}} &=&
\displaystyle{\int d^4x\, \overline\psi\gamma^\nu\gamma^\mu[\partial_\mu
(A+iB\gamma_5)]\partial_\nu\epsilon,}
\\[2mm]
\displaystyle{\d\int d^4x\,\LL^{d=4}_{ \rm mass}} &=&
\displaystyle{-iM_4 \int d^4x\, \overline\psi\gamma^\mu(A+iB\gamma_5)
\partial_\mu\epsilon },
\end{array}
\eeq
which of course vanishes at zero temperature where $\epsilon$ is a
constant spinor.

We now proceed with the thermal mode expansion of the action \equ{acto}, into
which we insert the Matsubara developments. For bosons,  the mode expansion
follows from \equ{bd} and \equ{abz}:
\be
\Gamma(t,\xx)={1\over\sqrt{\b}}\sum_{n=-\infty}^{n=\infty} \Gamma_n(\xx) \
e^{i\o_n^B t}\,,
\qquad \Gamma_{-n}=\Gamma^*_n\,,\qquad\Gamma=A,B,F,G\,. \label{deva}
\ee
More care is needed for the thermal modes of fermionic fields. Since the mode
expansion in the time direction effectively reduces the space-time dimension to
three, the $d=4$ Majorana spinor $\psi$ must be rewritten  in terms of a single
$d=3$ (euclidean) spinor $\lambda$. In our  conventions, the expression is:
\be
 \psi={{\left (\matrix{
\l\cr
i\s^2\l^*\cr
}\right )}}\,,
\qquad\qquad
\psi^\dagger = { \left (\matrix{
\l^\dagger & -i\l^T \s^2 \cr
}\right )}\,,
\qquad
(\bar\psi = \psi^\dagger \g^0)\,.\label{devpp}
\ee
For the two-component fermions $\l$, $\l^T$, we hence set
\be
\l(t,\xx)={1\over\sqrt{\b}}\sum_{n=-\infty}^{n=\infty} \l_n(\xx) \ e^{i\o_n^F
t}\,,\qquad \l^T(t,\xx)={1\over\sqrt{\b}}\sum_{n=-\infty}^{n=\infty}
\l^T_n(\xx) \ e^{i\o_n^F t}\,,\label{devl}
\ee
while the conjugates $\l^*$, $\l^\dagger$ are developed similarly, with
opposite frequencies.  Inserting these expansions into the $T=0$ supersymmetric
action \equ{acto}, one gets straightforwardly the euclidean\footnote{The finite
temperature expansion uses periodic {\it imaginary} time and the relevant
quantity to analyze the theory is the euclidean action.} action at
temperature $T=1/\b$:
\be\ba{l}
S_\b = \displaystyle{{1\over\b}\int_0^\b\!\! dt\!\intt
\!\!\!\!\sum_{m,n=-\infty}^{+\infty}}\!\Biggl\{
\Bigl[
\half \pa^i A_m \pa_i A_n
+\half \pa^i B_m \pa_i B_n
+ \half (\o_n^B)^2(A_m A_n+B_m B_n)
\\[6mm]
\hspace{3cm}
-\displaystyle\half(F_m F_n+G_m G_n) +M_4(A_m F_n+B_m G_n)
\Bigr]
\ e^{i(\o_m^B+\o_n^B)t}
\\[4mm]
\hspace{3cm}
+ \displaystyle\half \left[\l_m^\dagger
(\s^i\pa_i -\o_n^F)\l_n \, e^{i(\o_n^F - \o_m^F)t}
+ M_4\lambda_m^\dagger i\sigma^2\lambda_n^* e^{-i(\omega_m^F+\omega_n^F)t}
\right] + {\rm h.c.}\Biggr\}.
\ea\label{momo}\ee
Integrating on $t$, we get a three-dimensional theory of the
$\xx$-depen\-dent field modes $A_n$, $B_n$, $F_n$, $G_n$, $\l_n$, $\l^T_n$,
$\l^*_n$ and $\l^\dagger_n$ given by the {\it Matsubara action}:
\be\ba{l}
\!\!\!S^{d=3} = \displaystyle{\intt \!\!\!\!\sum_{n=-\infty}^{+\infty}}
\Biggl\{\half \pa^i A_{-n} \pa_i A_n
+\half \pa^i B_{-n} \pa_i B_n
+\half (\o_n^B)^2(A_{-n} A_n+B_{-n} B_n)
\\[2mm]
\phantom{\!\!\!S^{d=3}_\b = \displaystyle{\intt
\!\!\!\!\sum_{n=-\infty}^{+\infty}}
\Biggl\{}
-\displaystyle\half(F_{-n} F_n+G_{-n} G_n)
+\displaystyle\half \left[\l_n^\dagger
(\s^i\pa_i - \o_n^F)\l_n  + {\rm h.c.}\right]
\\[2mm]
\phantom{\!\!\!S^{d=3}_\b = \displaystyle{\intt
\!\!\!\!\sum_{n=-\infty}^{+\infty}}
\Biggl\{}
+M_4(A_{-n} F_n+B_{-n} G_n)+\displaystyle{M_4\over 2}\left[\l_{-n-1}^\dagger
i\s^2\l^*_n  + {\rm h.c.}\right]
\Biggl\}
\,.
\ea\label{mimi}\ee
The equations of motion are given here by the thermal modes of the $T=0$
equations \equ{eqmto},
\be
F_n=M_4\,A_n\,, \qquad\qquad G_n=M_4\,B_n\,.
\label{eqm}
\ee
The fields $A,B,F,G$ being real, we have for their thermal modes the relations
\be
A_{-n}=A^*_n\,,\qquad
B_{-n}=B^*_n\,,\qquad
F_{-n}=F^*_n\,,\qquad
G_{-n}=G^*_n\,.\label{rabfg}
\ee
Upon using the equations of motion \equ{eqm} and replacing \equ{rabfg}, we get
for the thermal expansion of the $d=4$, $T=0$ supersymmetric action
\equ{acto} on  $\real^3$ the euclidean expression
\be\ba{l}
S^{d=3} =\displaystyle{\intt \sum_{n=-\infty}^{+\infty}}
\Biggl\{
\displaystyle\half \pa^i A^*_{n} \pa_i A_n
+\displaystyle\half \pa^i B^*_{n} \pa_i B_n
+\displaystyle\half(M^B_{3,n})^2\,(A^*_{n} A_n+B^*_{n} B_n)
\\[2mm]
\phantom{S^{d=3} =\displaystyle{\intt \sum_{n=-\infty}^{+\infty}}
\Biggl\{}
+\displaystyle\half\left[
\l_n^\dagger (\s^i\pa_i -\o_n^F)\l_n
+ M_4  \l_{-n-1}^\dagger i\s^2\l^*_n\right] +{\rm h.c.}
\Biggr\}\,,
\ea\label{mumu}\ee
where $(M^B_{3,n})^2$ stands for the (squared) thermal mass of the $n$-th
$d=3$ bosonic mode,
\be
(M^B_{3,n})^2  = M_4^2 + (\o_n^B)^2, \qquad  (\o_n^B)^2={4\pi^2 n^2\over
\b^2}\
,\qquad n=-\infty, ... ,+\infty\,.\label{3bm}
\ee
For fermions, since $\omega_{-n-1}^F=-\omega_n^F$, $\lambda_n$ and
$\lambda_{-n-1}^\dagger$ are associated with the same time-dependent
phase. The mass matrix in \equ{mumu} writes then:
\beq
{\cal L}_{m,fermions}^{d=3} = {1\over2}\sum_n
\left( \lambda^\dagger_n \quad \lambda_{-n-1}\right)
\left( \begin{array}{cc} \omega_n^F & M_4 \\
M_4 & -\omega_n^F  \end{array}\right)
\left( \begin{array}{c} \lambda_n \\ \lambda^\dagger_{-n-1}
\end{array} \right)
+ {\rm h.c.}
\eeq
This mass matrix has two opposite eigenvalues $\pm M_{3,n}^F$,  verifying
the relation for the (squared) thermal mass of the $n$-th $d=3$ fermionic
mode
\be
(M^F_{3,n})^2  = M_4^2 + (\o_n^F)^2\,,\qquad  (\o_n^F)^2={\pi^2
(2n+1)^2 \over
\b^2}\,, \qquad n=-\infty, ... ,+\infty\,,\label{3fm}
\ee
as expected. The eigenstates are linear combinations of $\lambda_n$ and
$\lambda^\dagger_{-n-1}$. From eqs. \equ{3bm}, \equ{3fm}, it is clear
that thermal effects lift the mass degeneracy characteristic of $T=0$
supersymmetry.

In Section \ref{ttsstt}, we have shown that component transformations under
thermal supersymmetry have the same form as at $T=0$, but with the space-time
constant supersymmetry parameter $\epsilon$ replaced by the
thermal, time-dependent and antiperiodic quantity $\widehat\epsilon$.
This allows us  to identify immediately the thermal
version of the transformations \equ{ss2}:
\be\ba{rclcrcl}
\widehat\d A
\=
 \widehat{\bar\epsilon} \psi\,,
&&
\widehat\d B
\=
  i\widehat{\bar\epsilon}\g_5\psi
\,,\label{ss1}\\[3mm]
\widehat\d F
\=
 i\widehat{\bar\epsilon}\g^\m (\pa_\m \psi)\,,
&&
\widehat\d G
\=
  -\widehat{\bar\epsilon}\g_5 \g^\m(\pa_\m \psi)
\,,\\[3mm]
\widehat\d\psi
\=
 -[i\g^\m(\pa_\m(A+iB\g_5))+F+iG\g_5]\widehat\epsilon\,,
&&
\\[2mm]
\widehat\d\bar\psi
\=
 -\widehat{\bar\epsilon} [i\g^\m(\pa_\m(-A+iB\g_5))+F+iG\g_5]\,.&&
\ea\label{tss2}\ee
These expressions can be easily translated into transformations of the
Matsubara modes. The thermal supersymmetry
parameter is to be expressed in terms
of a three-dimensional two-component spinor $e(t)$  as
\be
\widehat\epsilon(t)={\left (\matrix{
\widehat e(t)\cr
i\s^2 \widehat e^*(t)\cr
}\right )}\,,\qquad\qquad
\widehat\epsilon^\dagger(t) = \left (\matrix{
\widehat e^\dagger(t) & -i\widehat e^T(t) \s^2 \cr
}\right )\,,\label{devee}
\ee
and  $\widehat e$, $\widehat e^T$, $\widehat e^*$ and $\widehat e^\dagger$ are
to be expanded thermally. Unlike  in eqs. \equ{devl}, the two-component
fermions here only depend on time. They therefore develop into frequency sums
with {\it constant} thermal modes $e_n$, $e^T_n$, $e^*_n$ and $e^\dagger_n$, as
\be
\widehat e(t)={1\over\sqrt{\b}}\sum_{n=-\infty}^{n=\infty}
e_n \ e^{i\o_n^F t}\,,
\qquad\qquad
\widehat e^T(t)={1\over\sqrt{\b}}\sum_{n=-\infty}^{n=\infty}
e^T_n \ e^{i\o_n^F t}
\,,\label{edevl}
\ee
and similarly for $e^*_n$ and $e^\dagger_n$, with opposite frequencies.
Inserting the mode expansions in eqs. (\ref{ss1}), the
transformations of bosonic Matsubara modes write
\beq
\d A_k
\={1\over\sqrt{\b}}
\sum_m\left(  ie^\dagger_{-k-m-1}\s^2\l^*_m -i
e^T_{k-m-1}\s^2\l_m\right)\,,\label{k1}\\[2mm]
\d B_k
\={1\over\sqrt{\b}}
\sum_m\left( e^\dagger_{-k-m-1}\s^2\l^*_m + e^T_{k-m-1}\s^2\l_m\right)\,
,\label{k2}\\[2mm]
\d F_k
\={1\over\sqrt{\b}}
\sum_m\left( -ie^\dagger_{m-k} [(\s^i\pa_i - i\o_m^F) \l_m] -i
e^T_{m+k} [(\s^{i\,T}\pa_i + i\o_m^F) \l^*_m]\right)\,,\label{k3}\\[2mm]
\d G_k
\={1\over\sqrt{\b}}
\sum_m\left( -e^\dagger_{m-k} [(\s^i\pa_i - i\o_m^F) \l_m] +
e^T_{m+k} [(\s^{i\,T}\pa_i + i\o_m^F) \l^*_m]\right)\,,\label{k4}
\eeq
while for fermionic Matsubara modes, upon introducing for the scalars $z$ and
$f$, eq. \equ{abz}, the notations $\zeta_m=\sqrt{2}\,z_m$,
$\varphi_m=\sqrt{2}\,f_m$, with
\be
\zeta_m = A_m+iB_m\,,\quad \widetilde\zeta_m = A_m-iB_m\,,\quad
\varphi_m = F_m+iG_m\,,\quad \widetilde\varphi_m = F_m-iG_m\,,\label{nott}
\ee
we get
\beq
\d \l_k
\={1\over\sqrt{\b}}
\sum_m\left(  [(\s^i\pa_i + i\o_m^B)\widetilde\zeta_m] \s^2 e^*_{m-k-1} -
e_{k-m} \varphi_m \right)
\,,\label{k5}\\[2mm]
\d \l^*_k
\={1\over\sqrt{\b}}
\sum_m\left( \s^2 [(\s^i\pa_i - i\o_m^B)\zeta_m]  e_{-k-m-1} -
e^*_{k+m} \widetilde\varphi_m  \right)
\,,\label{k6}\\[2mm]
\d \l^T_k
\={1\over\sqrt{\b}}
\sum_m\left( e^\dagger_{m-k-1} [(\s^i\pa_i - i\o_m^B)\widetilde\zeta_m] \s^2
 - e^T_{k-m} \varphi_m \right)
\,,\label{k7}\\[2mm]
\d \l^\dagger_k
\={1\over\sqrt{\b}}
\sum_m\left( e^T_{-k-m-1} \s^2 [(\s^i\pa_i + i\o_m^B)\zeta_m]
- e^\dagger_{k+m} \widetilde\varphi_m  \right)
\,.\label{k8}
\eeq
As the parameter $\widehat\epsilon$ has been expanded in non-trivial
Matsubara modes, these transformations mix in general boson and fermion
modes with different levels $n$, in contrast to the simple dimensional
reduction case where $\epsilon$ would be a zero-mode constant spinor.

In analogy with the zero-temperature transformation (\ref{transf1}),
the thermal action will have the following non trivial variation
under thermal supersymmetry (\ref{ss1}):
\beq
\label{transf2}
\begin{array}{rcl}
\displaystyle{\widehat\d\int d^4x\,\LL^{d=4}_{\rm kin}} &=&
\displaystyle{\int d^4x\, \psi^\dagger\gamma^\mu\partial_\mu
(A+iB\gamma_5)\partial_0\widehat\epsilon,}
\\[2mm]
\displaystyle{\widehat\d\int d^4x\,\LL^{d=4}_{ \rm mass}} &=&
\displaystyle{-iM_4 \int d^4x\, \psi^\dagger(A+iB\gamma_5)
\partial_0\widehat\epsilon },
\end{array}
\eeq
where $\partial_0\widehat\epsilon$ does not vanish.
In these expressions, it is understood that a rotation to imaginary (euclidean)
time is performed, and that time is integrated over the interval
$[0,\beta]$ only. Again, inserting the Matsubara mode expansions leads to
\beq
\label{Stransf}
\begin{array}{rcl}
\displaystyle{\widehat\d\int d^4x\,\LL^{d=4}_{\rm kin}} &=&
\displaystyle{{i\over\sqrt\beta}\int d^3x\, \sum_{m,n}
\omega^F_n\lambda^\dagger_{m-n-1}(\omega^B_m + i\vec\sigma\cdot\vec\nabla)
(A_m -iB_m )\sigma^2 e^*_n +{\rm h.c.},}
\\[3mm]
\displaystyle{\widehat\d\int d^4x\,\LL^{d=4}_{ \rm mass}} &=&
\displaystyle{{M_4\over\sqrt\beta}\int d^3x\, \sum_{m,n}
\omega_{n}^F(A_m-iB_m)\lambda^\dagger_{n+m}e_{n} + {\rm h.c.}}
\end{array}
\eeq
Clearly, neither the kinetic action nor the mass action are
invariant under the thermal supersymmetry transformations.
However, the variations $\d S^{d=4}_{\rm kin}$ and
$\d S^{d=4}_{\rm mass}$ vanish separately in the $T\rightarrow 0$ limit, as
expected. The variation of the total action is a combination of
two terms proportional\footnote{The prefactor $\beta^{-1/2}$
is a normalization of the mode expansion which disappears in the
$T\rightarrow 0$ limit.} to $\o_n^F \sim
T$.

\section{Conclusions}\label{concl}

Our conclusions are twofold. Firstly, superspace can be modified to satisfy
the constraints imposed by thermal effects and statistics.
In particular, superspace
Green's functions verifying the KMS conditions for bosons and fermions can
be written. The modified thermal superspace admits a super-Poincar\'e
algebra. It should be noted that this result is closely similar to the
Lorentz covariant formulation of finite temperature field theory \cite{weldon}.
The algebras of space-time or superspace transformations are essentially
local, they generate infinitesimal transformations
and are therefore not affected
by global conditions like periodicity or antiperiodicity along the
time/temperature circle.

Secondly, in contrast to the case of Lorentz
symmetry, the different statistics strongly affect realizations of
supersymmetry
on multiplets of fields at finite temperature. Thermal supersymmetry
of superfields is broken
because of the temperature-dependent constraints we impose on the
Grassmann coordinates of thermal superspace. These constraints
imply a covariantization of the $T=0$ superspace operators with respect
to temperature, which in turn indicates that a simple superfield will
not be sufficient to represent the thermal version of the
super-Poincar\'e algebra.

The discussion of the KMS conditions provides thermal superspace with a more
formal background. The requirement of antiperiodicity of the Grassmann
coordinates is essential both in proving a KMS condition at the
level of superfields, and in showing that the latter implies the correct
bosonic and fermionic boundary conditions for the superfield
components\footnote{A formulation of the KMS conditions at the superfield level
has been attempted in \cite{fuchs}. In that work, the superspace Grassmann
coordinates are taken constant and therefore the naive superfield KMS
conditions are seen not to hold. They are then reformulated in a
``superthermal" approach inspired from \cite{vh}.}. Notice that Green's
functions involving general superfields can be treated following the
method used here for chiral superfields only.

Signs of thermal supersymmetry breaking are seen only after the boundary
conditions that characterize thermal fields have been implemented. These
boundary conditions can be formulated in various equivalent ways, either in
terms of KMS conditions, or as periodicity and antiperiodicity requirements
on the fields, or equivalently upon expanding these fields thermally.
Irrespectively of the form in which they are implemented, the boundary
conditions carry information on the behaviour of thermal fields at distant
regions of space-time, and are in this sense of {\it global} nature. They
induce a strong differentiation between bosons and
fermions, as bosonic fields are periodic in imaginary time, while fermionic
ones are antiperiodic\footnote{Previous studies of thermal
supersymmetry breaking, considered either as explicit or as spontaneous, \eg
\cite{dk,ggs,fujikawa,boyan,gs,bo}, or of thermal Lorentz breaking
\cite{aoyama,ojima}, have  been conducted at the level of thermal fields/states
with this global periodicity/antiperiodicity distinction.} and generate
obstructions  -- in terms of lifting of the $T=0$ mass degeneracy and of
non-invariance of the Matsubara action -- when  trying to realize the thermal
super-Poincar\'e algebra on systems of thermal fields. But thermal superspace
can be used to analyze these obstructions.

\vspace{1cm}
\noindent{\bf Acknowledgements}\\[2mm]
The authors acknowledge collaboration with C. Manuel in early stages of
this work.

%***********************************************************
\end{document}